\documentclass[useAMS,usenatbib]{mn2e}

\usepackage{times}
\usepackage{graphicx}

\newcommand{\singlecolsize}{0.48}

\newcommand{\Msun}{\mathrm{M}_{\odot}}

\newcommand{\sqdeg}{{\rm deg}^2}

\newcommand{\hunits}{{\rm\,km\,s^{-1}\,Mpc^{-1}}}

\newcommand{\rmodel}{r_{\rm model}}
\newcommand{\dsgprof}{\Delta_{\rm sg,prof}}
\newcommand{\dsgjk}{\Delta_{\rm sg,jk,gi}}
\newcommand{\dsgyk}{\Delta_{\rm sg,yk,gz}}
\newcommand{\reff}{r_{50,\mathrm{maj}}}

\newcommand{\midtilde}{\raisebox{-0.25\baselineskip}{\textasciitilde}}

\newcommand{\newtext}[1]{{#1}}  

\title[Compact galaxies and the size-mass galaxy distribution]
      {Compact galaxies and the size-mass galaxy distribution from a colour-selected sample
      at $0.04 < z < 0.15$ supplemented by $ugrizYJHK$ photometric redshifts}

\author[I.~K.~Baldry et al.]
{{\parbox{\textwidth}{\raggedright 
Ivan~K.~Baldry, 
Tricia~Sullivan, 
Raffaele~Rani,
Sebastian~Turner
}}\\
\vspace{0.4cm}\\
{\parbox{\textwidth}{\raggedright 
Astrophysics Research Institute, Liverpool John Moores University, IC2, Liverpool Science Park, 146 Brownlow Hill, Liverpool, L3 5RF, UK
\\
}}}

\begin{document}

\date{Accepted by MNRAS, 2020 October 22}

\pagerange{\pageref{firstpage}--\pageref{lastpage}} \pubyear{2020}

\maketitle

\label{firstpage}

\begin{abstract}
  The size-mass galaxy distribution is a key diagnostic for galaxy evolution.
  Massive compact galaxies are potential surviving relics of a high-redshift phase of star formation.
  Some of these could be nearly unresolved in SDSS imaging and thus not included in galaxy samples. 
  To overcome this, a sample was selected from the combination of SDSS and UKIDSS photometry to $r<17.8$.
  This was done using colour-colour selection, and then by obtaining accurate photometric redshifts (photo-z)
  using scaled flux matching (SFM). Compared to spectroscopic redshifts (spec-z), SFM obtained
  a 1-sigma scatter of 0.0125 with only 0.3\% outliers ($|\Delta\ln(1+z)|>0.06$). 
  A sample of 163\,186 galaxies was obtained with $0.04<z<0.15$ over $2300\,\sqdeg$ using a combination of spec-z and photo-z.
  Following Barro et al., $\log\Sigma_{1.5}{}={}\log{}M_{*}-1.5\log\reff$ was used to define compactness.
  The spectroscopic completeness was 76\% for compact galaxies ($\log\Sigma_{1.5}>10.5$) compared to 92\% for normal-size galaxies.
  This difference is primarily attributed to SDSS `fibre collisions' and not the completeness of the main galaxy sample selection.
  Using environmental overdensities, this confirms that compact quiescent galaxies are significantly 
  more likely to be found in high-density environments compared to normal-size galaxies. 
  By comparison with a high-redshift sample from 3D-HST, $\log\Sigma_{1.5}$ distribution
  functions show significant evolution, with this being a compelling way to compare with simulations such as EAGLE.
  The number density of compact quiescent galaxies drops by
  a factor of about 30 from $z\sim{}2$ to $\log{}(n/\mathrm{Mpc}^{-3}){}={}-5.3\pm{}0.4$ in the SDSS-UKIDSS sample.
  The uncertainty is dominated by the steep cut off in $\log\Sigma_{1.5}$,
  which is demonstrated conclusively using this complete sample. 
\end{abstract}


\section{Introduction}
\label{sec:intro}

The galaxy population of the local Universe is very different to
its ancestral population ten billion years ago at $z \sim 2$. 
One of the most striking changes is the transformation of the radially small
massive galaxies seen at high redshift to the larger and more diffuse
red galaxies seen today. Multiple observations indicate that a
high proportion of galaxies at $z > 2$ are already `red and dead' as
well as unusually small for their mass relative to local red galaxies 
(e.g., \citealt{daddi05,trujillo07}). 
A quintessential high-redshift quiescent galaxy, or `red nugget', 
has a stellar mass ($M_{*}$) of $\sim 10^{11} \Msun$ and 
a half-light radius of  $\sim 1\,\mathrm{kpc}$ \citep{buitrago08,vandokkum08,damjanov09}. 

Considering the demographics in more detail,
at $z\sim2$ almost half of the massive ($M_{*} > 10^{10} \Msun$) galaxy population are both quiescent 
(e.g.\ \citealt{cimatti04,kriek06}) and radially smaller than quiescent galaxies locally 
\citep{daddi05,trujillo07,longhetti07,buitrago08,barro13}.
The red nuggets at $z \sim 2.3$ measure 5--6 times smaller than the median
size of low-redshift samples \citep{shen03,lange15} for the same mass \citep{vandokkum08}. 
These observations have posed a significant challenge for theories of galaxy evolution. 

In a hierarchical structure formation scenario,  
red nuggets should evolve into the most massive 
ellipticals in the local Universe (e.g., \citealt{trujillo07}). 
An early proposed mechanism for this evolution involved an 
adiabatic expansion of stellar systems in response to a 
significant mass loss from their inner regions. This mass expulsion is 
thought to be driven by means of stellar winds or/and 
quasars and to increase the effective radius as $R_e \propto M^{-1}$ 
\citep{fan08,damjanov09,fan10,RFG11}. 
However, the amount of the stellar mass loss is 
too small to explain the size evolution \citep{damjanov09}.
Compact massive galaxies are already quiescent in terms of star-formation 
activity at high $z$ and they do not seem to possess 
sufficient gas to increase their stellar component significantly 
\citep{bezanson09}. 

A natural hierarchical explanation for this radial growth would involve
major mergers ($\sim 1$:1 mass ratio) building the red sequence from the
high-redshift compacts to the local red giant ellipticals. 
However, major mergers result in equal growth of both mass and radius \citep{naab09}; 
they shift galaxies along the high-redshift mass-radius relation rather than 
towards the low-redshift relation 
(see also \citealt{boylan-kolchin06}). 
Major mergers are effectively ruled out of consideration for this reason, 
and because `wet' (star-forming) mergers would leave a
signature in the stellar density profiles of galaxies that is not seen
\citep{szomoru12}.

The current view for explaining the size-mass evolution 
is that a series of minor mergers together with accretion of
a stellar envelope have combined to build either local elliptical
galaxies or massive spiral/S0 galaxies 
\citep{hopkins10,sonnenfeld14,vandokkum14,graham15,buitrago17}. 
Progenitor bias has also been
invoked in explaining the apparent size evolution of the red galaxy 
population as a class (e.g.\ \citealt{szomoru12,carollo13}). 
In fact, as noted by \citet{barro13}, both paths must operate: 
an early formation of massive compact galaxies that become quenched and then 
grow via minor mergers \citep{naab09,hilz12}, 
and a late-arrival path in which larger star-forming galaxies build mass 
and then form extended quenched galaxies \citep{barro13}. 

It has long been recognised that local massive ellipticals generally have old stellar populations 
with little recent star formation. 
However, they could still have undergone significant evolution due to mergers unless they are compact. 
The stochastic nature of merging processes suggests that a non-negligible number of 
quiescent massive compact galaxies at each redshift between 2 and 0 should be unaltered, 
old and compact `relics' \citep{trujillo14}. 
Quantifying the number density of these massive compact relics in the present-day universe 
is thus an important constraint on galaxy formation models \citep{QT13,ferre-mateu17}, 
in addition to relics being local analogs of the high-redshift red nuggets
\citep{saulder15,yildirim17,martin-navarro19}. 

\citet{valentinuzzi10local} searched for and found a population of compact galaxies in
local $0.04 < z < 0.07$ clusters using mass and surface mass density lower limits.
\citet{poggianti13} showed that the fraction of these `superdense' galaxies was 3--4 times
higher in groups with high velocity dispersion ($> 500$\,km/s) compared to the field.
This means that it requires large volumes using blind surveys in order to sample
compact galaxies in high-density environments. 
\citet{saulder15} searched the Sloan Digital Sky Survey (SDSS) for compact quiescent
galaxies, using a lower limit on galaxy velocity dispersion ($> 300$\,km/s)
in addition to an upper limit on size ($\la 2$\,kpc), finding only 76 galaxies at $0.05 < z < 0.3$.
Even these are not as extreme as the high redshift red nuggets.
Using deeper imaging in the Galaxy And Mass Assembly (GAMA) survey regions, \citet{buitrago18}
found 22 massive compact galaxies ($\log M_{*} > 10.9$) with number density 
$\sim 10^{-6}\,\mathrm{cMpc}^{-3}$ at $z<0.3$.
Various authors have quantified the evolution in the number density of
compact galaxies (e.g.\ \citealt{barro13,cassata13,vandokkum15,tortora16,charbonnier17}),
for $\log M_{*} > 10$ or higher limits, for quiescent/star-forming galaxies,
and with various definitions of compactness (e.g.\ \citealt{gargiulo16,lu19}).
The number density of compact quiescent galaxies peaks around $z \sim 1$--2.

There is some concern that local galaxy surveys could be missing extremely compact galaxies
due to mis-identification with stars in the input catalogue. For example, 
for the SDSS main galaxy sample, $r<17.77$\,mag, a profile separator is used for galaxy selection
\citep{strauss02}. Without this, there would be about ten times as many stars as
galaxies for this magnitude limit. 
\citet{liske06} spectroscopically identified all sources over $1.14\,\sqdeg$ to $B<20$\,mag
for the Millenium Galaxy Catalogue (MGC). They estimated that about 1\% of galaxies in the MGC
input catalogue were mis-identified as stars.
The volume probed in the $\sim1\,\sqdeg$ region, however, was not sufficient to
search for massive compact galaxies.
\citet{taylor10} considered the selection of galaxies in SDSS, which includes
cuts against saturation, spectroscopic fiber crosstalk, and the concentration of light. 
All of these characteristics predispose compact galaxies to exclusion by the SDSS automated data
pipeline for spectroscopic selection.
\citeauthor{taylor10} concluded that the SDSS completeness
should be $\ga 75$\% for the types of compact galaxies 
seen at high redshift but maybe as low as $\sim 20$\% for the smallest galaxies. 

The aim of this paper is to determine the local number densities of compact galaxies using a 
complete sample. This is enabled by star-galaxy separation using colours and photometric
redshifts based on 9-band photometry from the combination of SDSS \citep{york00} and
the UKIRT Infrared Deep Sky Survey (UKIDSS, \citealt{lawrence07}). 
The sample selection and photometric redshifts are described in
\S~\ref{sec:sample-selection} and \S~\ref{sec:photo-z}.
The results of the size-mass relations and distributions are presented and discussed in
\S~\ref{sec:size-mass}, and a summary is presented in \S~\ref{sec:summary}.
\newtext{Size tests and images of compact galaxies are presented in the Appendix.} 
We assume a flat $\Lambda$CDM cosmology with $H_0=70\hunits$ and $\Omega_{m,0}=0.3$.




\section{Sample selection}
\label{sec:sample-selection}

\subsection{Data}
\label{sec:data}

Sources were selected from the SDSS DR14 database with $r<17.8$ ($r$-band
extinction-corrected Petrosian magnitude) covering five separate sky areas as
shown in Table~\ref{tab:sky-areas} (from tables \textsc{PhotoPrimary} and,
where available, \textsc{SpecObj}).  This approximately matched the UKIDSS
Large Area Survey (LAS) sky coverage.  No restrictions were applied based on photometric flags or
star-galaxy separation.  This produced a catalog of 3.38 million sources
covering $2664\,\sqdeg$.

\begin{table}
  \caption{Sky areas. RA and DEC limits for selection of sources
    from catalogs. The areas were chosen to approximately match
    UKIDSS LAS coverage.}
  \label{tab:sky-areas}
\centerline{
\begin{tabular}{ccc} \hline
  RA range (deg.) & DEC range (deg.) & note \\ \hline
  125 to 238 & $-2$ to 15 & LAS main region \\
  114 to 128 &  ~18 to 30 & \\
  190 to 210 &  ~22 to 36 & \\
  240 to 250 &  ~22 to 32 & \\
  $>309.2$ or $<58.6$ & $-1.26$ to 1.26 & Stripe 82 \\ \hline
\end{tabular}
}
\end{table}

Sources were selected from the UKIDSS database \textsc{lasYJHKsource} table,
which provides combined data for the LAS survey.  The magnitude limits were
$Y < 18.4$ or $J < 18.1$ with magnitude types petromag, apermag3 or apermag4.
These limits were a magnitude fainter than expected for any typical type of
galaxy given the SDSS $r$-band limit.  The aim was to be inclusive at this
stage.  This produced a catalog of 15.28 million sources.

Sources were selected from the GAIA DR2 \textsc{gaia\_source} table over the
areas defined in Table~\ref{tab:sky-areas}.  This produced a catalog 12.02
million sources, of which, 687\,471 have a GAIA $G$-band magnitude brighter
than 15.

\subsection{Masking radii around GAIA sources}
\label{sec:gaia-masking}

The SDSS main galaxy sample \citep{strauss02}
excluded sources for which the saturated flag was
set. This was to remove artifacts that are caused by light either diffracted
or scattered from bright stars. Without resorting to this flag, it is instead
possible to remove these types of artifacts by masking the sky area around
selected stars. These stars are, of course, uncorrelated with the galaxy
distribution and there is no bias in trimming the catalogue using this method.

\begin{figure}
  \centerline{\includegraphics[width=\singlecolsize\textwidth]{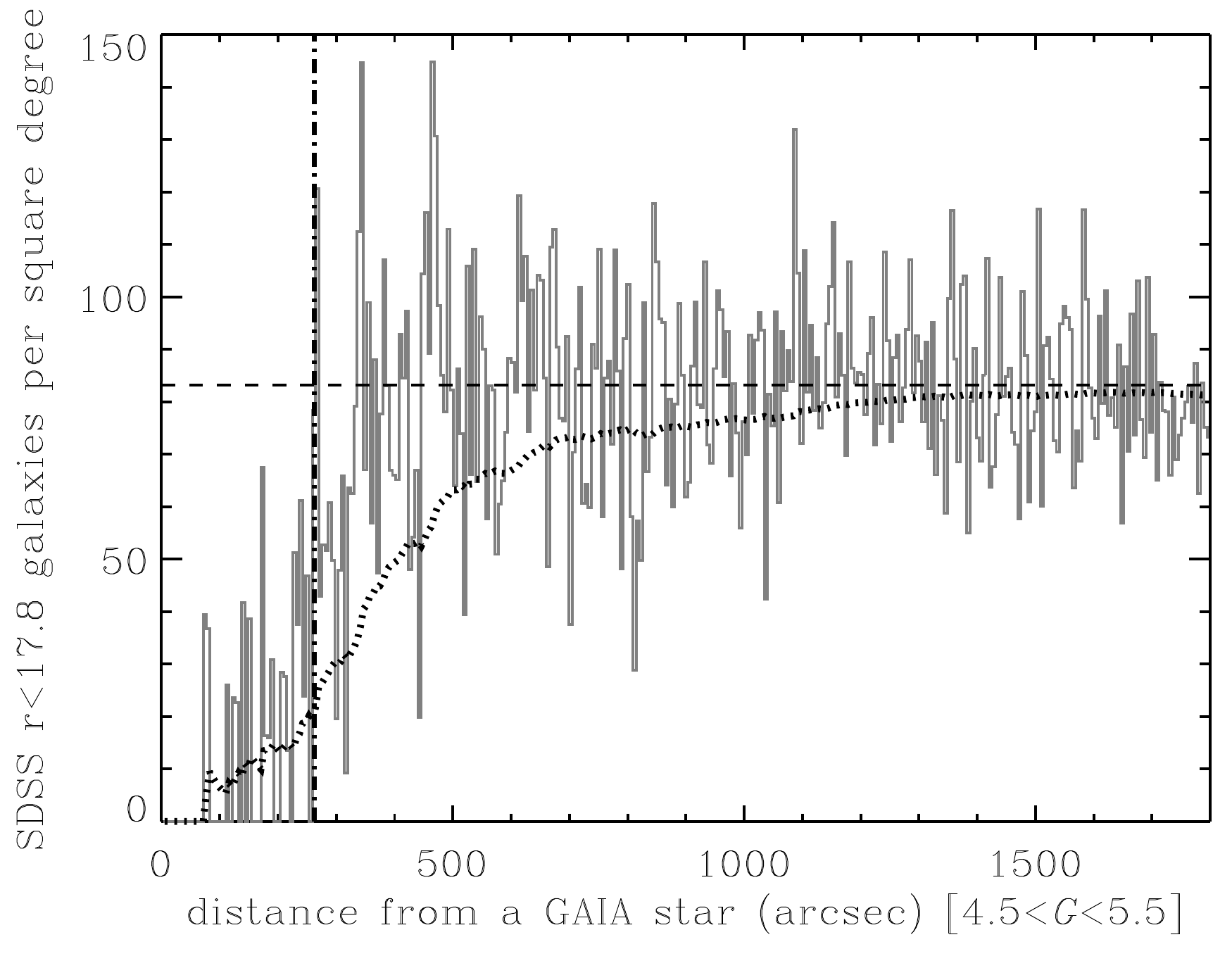}}
  \centerline{\includegraphics[width=\singlecolsize\textwidth]{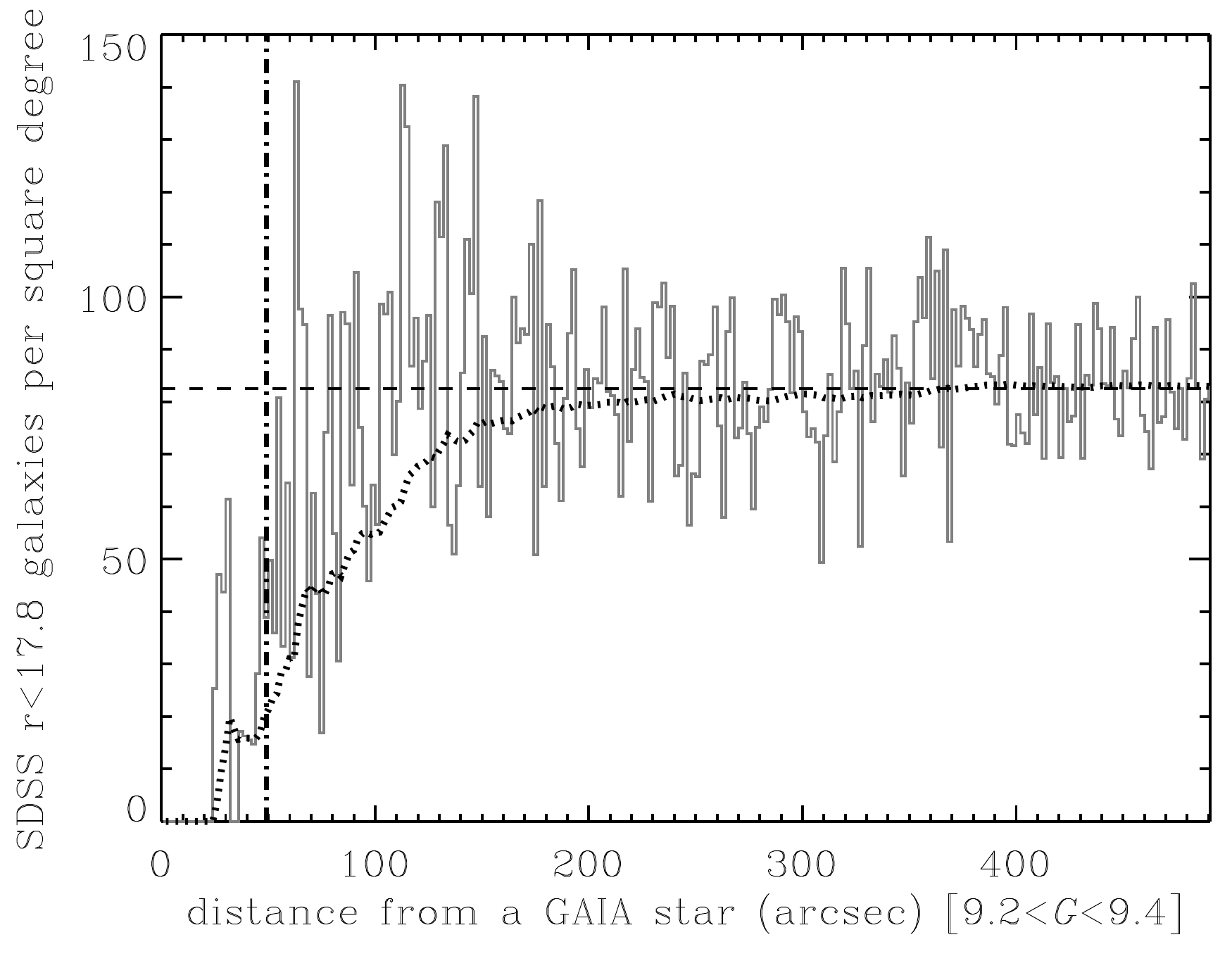}}
\caption{Examples of SDSS spectroscopically-confirmed galaxy sky densities
  around GAIA stars. 
  The grey solid-line shows the sky density histogram. 
  The dashed line shows the mean sky density away from the stars. 
  The dotted line shows the cumulative sky density (accounting for sky area).
  The vertical dash-dot line show the radius at which the cumulative sky
  density is 25\% of the mean.
  This corresponds to the empirical estimate of the exclusion radius.}
\label{fig:exclusion-examples}
\end{figure}

\begin{figure}
  \centerline{\includegraphics[width=\singlecolsize\textwidth]{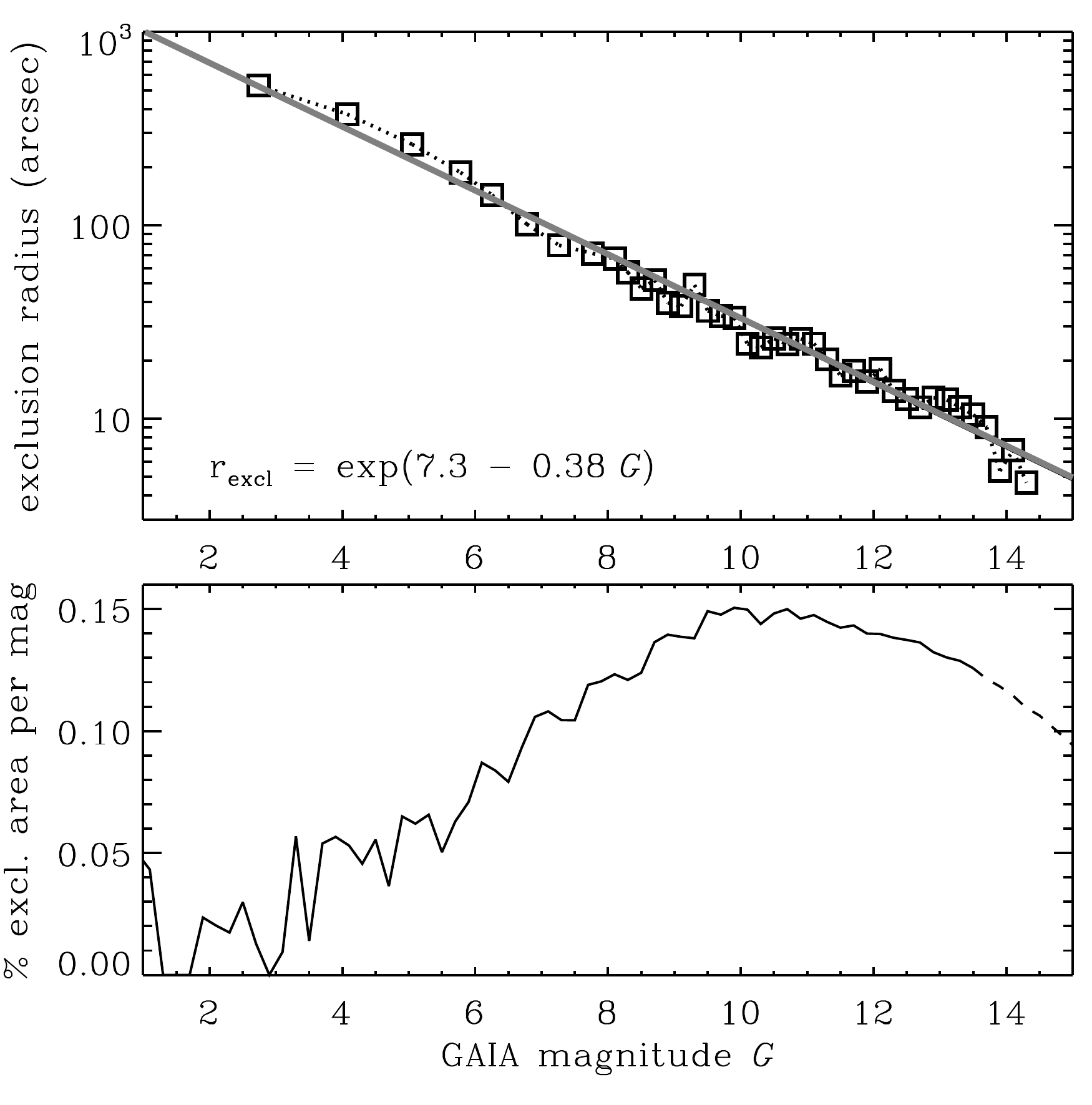}}
\caption{Top panel: exclusion radius versus GAIA $G$ magnitude.
  The squares and dotted line represent the results from the empirical
  determination for each bin (examples in Fig.~\ref{fig:exclusion-examples}).
  The solid grey line shows a fit to the these points, with
  the parameters and equation annotated.
  Lower panel: percentage of excluded area including all stars
  and using exclusion-radius values from the fit.
  The solid line shows the magnitude range used for this project,
  while the dashed line shows a small extension to 15th magnitude.}
\label{fig:exclusion-radius-fit}
\end{figure}

The exclusion radius around GAIA stars was empirically determined using the
spectroscopically-confirmed galaxy sample from SDSS.  To do this, the sky
density of confirmed galaxies around each star was measured in bins of
separation and GAIA magnitude of the star. Two examples of this are shown in
Figure~\ref{fig:exclusion-examples}.
The radius at which the sky density (within the radius) is 25\% of the 
expected sky density was determined. This was chosen as visually this is
the radius at which the impact of the stellar stray light becomes
negligible. Figure~\ref{fig:exclusion-radius-fit} shows the resulting
exclusion radii versus the GAIA magnitude. 


\subsection{Trimming the area}
\label{sec:trim-area}

The sample selection requires UKIDSS data and since the initial query did not
take account of the exact coverage, significant adjustment to the area was
needed. In particular, there is a requirement for $Y$, $J$ and $K$ data, and
the UKIDSS coverage was defined with the criteria that valid photometry exists
in all these bands. At the same time, there are genuine SDSS galaxies without a
UKIDSS catalogue match and these should not be rejected.  Therefore the SDSS
sample was trimmed by area and not by match criteria.

The SDSS master sample was trimmed in area using the following:
\begin{itemize}
\item Seven polygons were defined where there was no UKIDSS coverage
  and sources within these polygons were removed. 
\item A $6'\times6'$ grid was defined and sources in grid areas without any 
  UKIDSS coverage were removed.
\item One polygon was defined where there was no SDSS
  spectroscopy and sources within this polygon were removed. 
\item Sources in areas with $g$-band Galactic extinction greater than 0.4
  were removed.
\item Sources in the masked areas around $G < 13.5$ GAIA stars 
  were removed (\S~\ref{sec:gaia-masking}). Note GAIA sources include 
  galaxies but no spectroscopically-confirmed SDSS galaxies had a GAIA
  aperture\footnote{GAIA $G$-band flux measurements use a
    rectangular aperture of
    $0.7'' \times 1.05''$ ($12 \times 18$ pixels)
    for sources brighter than 16th mag, and
    $0.7'' \times 0.7''$ for fainter sources \citep{gaia16}.}
  magnitude brighter than 14.5 in the sample. Therefore, 
  13.5 is a safe limit that even compact galaxies would not be excluded. 
\end{itemize}
This produced a catalog of 2.48 million sources covering $2300\,\sqdeg$. 

\subsection{Colour-colour galaxy selection}
\label{sec:colour-colour}

In order to avoid using a profile separator for star-galaxy separation, 
galaxies were selected using a combination of SDSS and UKIDSS colours. 
To do this, we considered two colour-colour diagrams: 
$J-K$ versus $g-i$ and $Y-K$ versus $g-z$. 
These show the cleanest separation between galaxies and stars,
while also not requiring matched-aperture or matched-profile photometry
between UKIDSS and SDSS. For UKIDSS, we used \textsc{apermag4}
\newtext{($2.8''$-diameter aperture)}
and for SDSS, we used \textsc{psf} and \textsc{Petro} magnitudes.
\newtext{The UKIDSS aperture was chosen to provide accurate colours
  for stars and galaxies, $2.8''$ is more than twice the typical seeing 
  of $<1''$ \citep{dye06}. The SDSS \textsc{psf} magnitudes provide
  the most precise colours for point sources whereas \textsc{Petro} magnitudes 
  provide the best unbiased colours for extended sources.}


The basic method is to obtain a quadratic fit to the stellar locus
of the UKIDSS colour as a function of the SDSS colour over a suitable
range, with clipping of the function beyond that range. 
The galaxy selection is then defined as galaxies with UKIDSS colour 
greater than 0.2 above the function value. 

The $J-K$ separator is given by 
\begin{equation}
 \dsgjk = J_{\rm AB} - K_{\rm AB} - f_{\rm locus}(g-i)
\label{eqn:jkgi}
\end{equation}
with the locus function as per \citet{baldry10}. 
For this paper, we determined the locus function for $Y-K$ versus 
$g-z$ colours using spectrosopcially-confirmed stars 
and with point-spread function (PSF) magnitudes for the SDSS colours. 
This additional separator is given by 
\begin{equation}
 \dsgyk = Y_{\rm AB} - K_{\rm AB} - f'_{\rm locus}(g-z)
\label{eqn:ykgz}
\end{equation}
where
\begin{equation}
 f'_{\rm locus}(x) = 
\begin{array}{l} 
  -0.8297 \\ -1.04 + 0.74 x - 0.13 x^2 \\ 0.0128  \end{array}
\mbox{~for~} 
\begin{array}{l} 
  x < 0.3 \\ 0.3 < x < 2.8 \\ x > 2.8
\end{array}
\label{eqn:f_locus}
\end{equation}

\begin{figure}
\centerline{\includegraphics[width=\singlecolsize\textwidth]{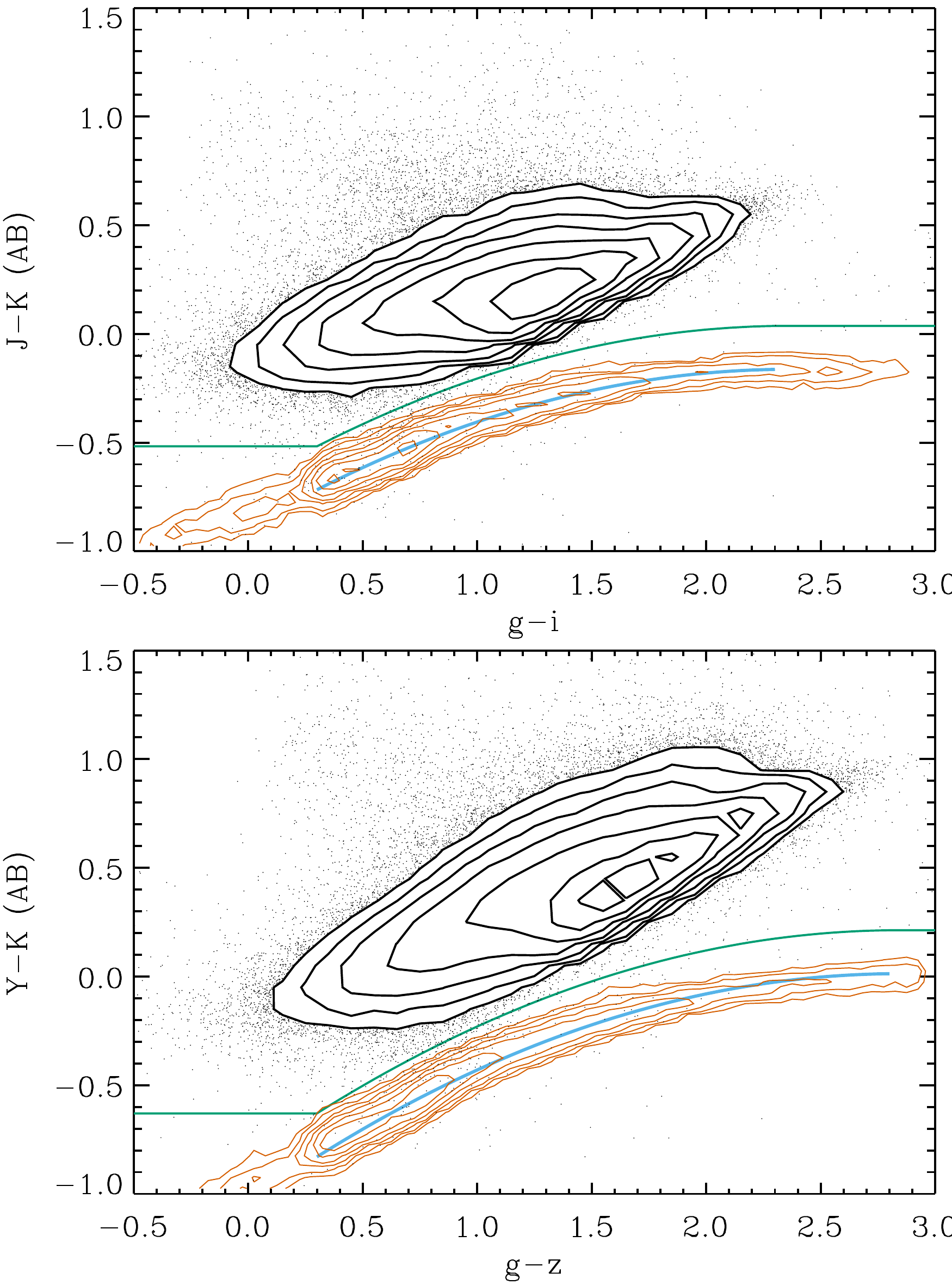}}
\caption{Colour-colour distributions for spectroscopically-confirmed
  galaxies and stars. The distributions are shown by contours and 
  points. The blue lines show the stellar locus fits. 
  The green lines show the selection boundaries for galaxies, 
  0.2 above the locus function.}
\label{fig:color-color}
\end{figure}

\begin{figure}
\centerline{\includegraphics[width=\singlecolsize\textwidth]{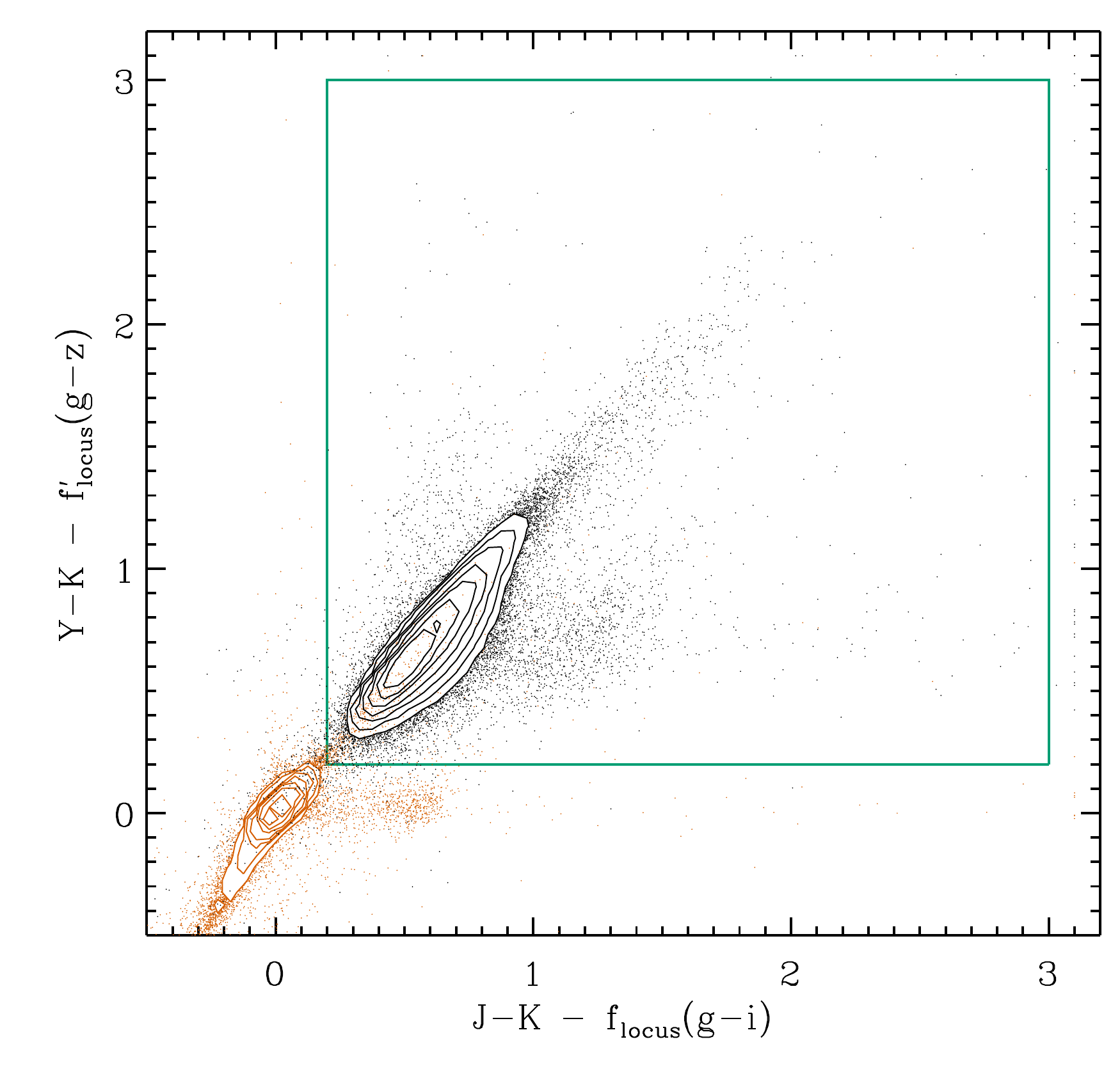}}
\caption{Boundaries in the $\Delta$ values used to select galaxies. 
  The distributions are shown for spectroscopically-confirmed galaxies
  (black contours and points) and stars (orange contours and points).}
\label{fig:delta-color}
\end{figure}

Figure~\ref{fig:color-color} shows the two colour-colour diagrams 
used to distinguish stars from galaxies. Note that when 
determining the $\Delta$ values for the selection, 
the bluer of SDSS PSF and SDSS Petrosian magnitudes are used for each source. 
This is in order to be inclusive toward galaxy selection 
(where the Petrosian colours could be different from the PSF colours). 
The colour-only galaxy selection is then given by 
\begin{equation}
  0.2 < \dsgjk < 3  \mbox{~~~\&~~~}  0.2 < \dsgyk < 3 
\end{equation}
with the high cuts used to exclude some cases of 
photometric measurement problems. 
Figure~\ref{fig:delta-color} shows these selection
boundaries with the distributions of $\dsgyk$
versus $\dsgjk$ for stars and galaxies. 

The colour-colour selection results in 225\,634 sources. 
Of these, 198\,123 have an associated reliable SDSS redshift, 
with 99.2\%, 0.3\% and 0.5\% classified as a galaxy, star and quasar, respectively
(a galaxy being defined to have $0.002<z<0.4$ or by spectral class from the SDSS pipeline). 
This demonstrates the high purity of this selection. 

\subsection{Additional selections}

The above selection is not complete because of the reliance on SDSS-UKIDSS matching,
and this can miss extended sources where the UKIDSS aperture position is offset or undefined.
In addition, the method of trimming the area to UKIDSS $Y$, $J$ and $K$ coverage is not perfect.
Additional sources with SDSS photometry but {\em no} UKIDSS match
were selected using the SDSS star-galaxy profile separator, 
\begin{equation}
  \dsgprof = \rmodel - r_{\rm psf} \mbox{~~~,}
\end{equation}
satisfying the following criteria: 
\begin{equation}
  \dsgprof > 0.5
  \mbox{~~~\&~~~ not \textsc{saturated} ~~~\&~~~}
  \mu_{r,50} < 24.5 \mbox{~~~,}
\end{equation}
where $\mu_{r,50}$ is the Petrosian half-light surface brightness. 
These criteria selected 15\,309 sources.

On top of the UKDSS-SDSS and SDSS-only selections,
4533 sources that were galaxy targets in DR7 
and 169 spectroscopically-confirmed galaxies,
but not otherwise selected, were included. 
Including all four selections,
this results in 245\,645 sources,
of which 86.1\% have redshifts from SDSS and 0.2\% from 2MRS \citep{huchra12}.
This leaves 13.7\% without spectroscopic redshifts (spec-z). 




\section{Photometric redshifts from scaled flux matching (SFM)}  
\label{sec:photo-z}

The above combined selection (\S~\ref{sec:sample-selection}) primarily
selects galaxies. However, 1.0\% of the sources are
spectroscopically-confirmed stars (kept in for testing purposes) and a
significant fraction of those without spec-z are stars. This is
because there are about 10 times as many stars as galaxies at these
magnitudes ($r < 17.8$), and finding any nearly unresolved galaxies is
like `looking for a needle in a haystack'.  To refine the search
further, we use a photometric redshift (photo-z) technique that identifies
sources with galaxy-like fluxes (beyond that already considered with
the colour-colour selection). In addition, redshifts are needed to
estimate luminosities and physical sizes.

Photo-z methods are generally divided into two classes:
template fitting (e.g.\ \citealt{babbedge04,BvDC08}) and empirical. 
When there are many sources with reliable spec-z covering the colour and magnitude space of
a sample, which is the case for this SDSS-UKIDSS sample, 
empirical methods should be more accurate \citep{Koo99}.
Here, we note that empirical methods can be further divided into: 
(a) predictive modelling, for example, using neural networks \citep{firth03}
or analytical functions \citep{connolly95,sedgwick19}; and
(b) photometric-property matching, typically using colours and a magnitude \citep{baldry06,beck16}.
In the latter method, a source's photo-z is obtained by the distribution
of spec-z for galaxies with similar properties. 
Here, we use scaled flux matching (SFM). This is similar to colour matching 
except that it works in linear flux space with the advantage that it can naturally 
deal with missing data, low S/N measurements \citep{sedgwick19conf} 
or, more generally, different errors in each band. 



\subsection{General method}

For each source $i$ and band $k$, 
the uncertainty in the flux is given by: 
\begin{equation}
  \sigma_{i,k}^2 = \sigma_{i,k,\mathrm{P}}^2 + (a_k f_{i,k})^2 
\end{equation}
where $\sigma_{i,k,\mathrm{P}}$ is the Poisson/counting or linear error, 
$a_k$ is the band-dependent fractional error, and 
$f_{i,k}$ is the flux. 
If the Poisson error is not well determined, then a band-dependent 
value could be used. 
The flux measurements and flux uncertainties are also corrected for 
Galactic extinction. 

For each source, the fluxes are matched to the matching-set galaxy ($j$) fluxes with 
chi-squared defined as: 
\begin{equation}
  \chi^2_{i,j} = \sum_k \frac{(f_{i,k} - n_{i,j} f_{j,k})^2}{\sigma_{i,k}^2}
\end{equation}
where $n_{i,j}$ is the usual best-fit normalization when scaling 
a model to fit data points with errors
(obtained from solving  $\mathrm{d}\chi^2 / \mathrm{d} n = 0$):
\begin{equation}
n_{i,j} = \frac {\sum_k f_{j,k} f_{i,k} / \sigma_{i,k}^2}
                {\sum_k f_{j,k}^2 / \sigma_{i,k}^2} \mbox{~~~.}
\end{equation}
Note that no uncertainty is applied to the matching-set fluxes 
in the calculaton of chi-squared.\footnote{Note for coding purposes in
  languages like IDL, R and \textsc{python}, the calculations can be performed
  using inbuilt matrix multiplation where appropriate rather than loops. This is known to be 
  signficantly faster.}

The reliability weight of the match is then given by: 
\begin{equation}
w_{i,j} = \exp( -\chi^2_{i,j} / 2) \, W_{j}
\end{equation}
where $W_{j}$ is any additional weight 
assigned to galaxy $j$ of the matching set. 
This could be, for example, to downweight
redshift bins that are well populated. 
Importantly, $w_{i,j}$ is set to zero where $i$ and $j$ refer to the same galaxy. 
This is so the photo-z of the 
matching-set galaxies are independent of their spec-z.

These weights could then be used to estimate
a probability distribution function (PDF) if desired (see \citealt{Wolf09}). 
In general, this method is accurate if the matching
set is large and the galaxies for which photo-z 
are desired are within the covered range
of spectral-energy distribution (SED) type and redshifts. A characterization 
of the PDF is then given by the weighted mean and standard deviation. 
The weighted mean of the redshift variable is given by:
\begin{equation}
  \zeta_{i,\mathrm{phot}} = 
  \frac{\sum_j w_{i,j} \zeta_{j,\mathrm{spec}}} {\sum_j w_{i,j}}
  \mbox{~~~where~~~} \zeta = \ln(1+z) \: .
\end{equation}
Noting that $\zeta$ (zeta) is the appropriate quantity to use when dealing with redshift measurements
and errors \citep{Baldry18zeta}.\footnote{To reduce computation time and to obtain an estimate 
of the weighted mean even if too many matching-set galaxies have similar weight, 
only a set number of the matching-set galaxies with the highest weights (lowest $\chi^2$) 
are included in the calculation of the weighted mean and nominal error. In this paper,
we used 2500. For comparison, \citet{beck16} used the best 100 matching-set galaxies with
all these given the same weight.}

The initial estimate of the uncertainty (nominal error) is given by: 
\begin{equation}
  \zeta_{i,\mathrm{err}}^2 
  = \left( \frac{\sum_j w_{i,j} \zeta_{j,\mathrm{spec}}^2}
         {\sum_j w_{i,j}} \: - \:  \zeta_{i,\mathrm{phot}}^2 \right)
    \frac{N_{i,\mathrm{eff}}}{N_{i,\mathrm{eff}} - 1} 
\label{eqn:nominal-error}
\end{equation}
which is the weighted standard deviation multiplied by a correction
factor to obtain the sample standard deviation.
The effective number of measurements for reliability weights 
is given by
\begin{equation}
  N_{i,\mathrm{eff}} =  \frac{ \left(\sum_j w_{i,j}\right)^2 }{ \sum_j w_{i,j}^2 } \mbox{~~~.}
\end{equation}

\subsection{Specific implementation}

SDSS \textsc{modelmag} values were converted to fluxes \citep{LGS99} 
in units of nanomaggies (log-flux zeropoint of 22.5\,mag, AB system; \citealt{blanton05nyuvagc}). 
\newtext{For photo-z measurements, \textsc{modelmag} fluxes were chosen because they provide 
  the highest S/N for point and extended sources.} 
UKIDSS \textsc{apermag4} (1.4$''$-radius aperture) magnitudes were converted to fluxes
in the same units 
(Vega to AB magnitude conversion was taken to be $+0.63,+0.94,+1.38,+1.90$ for $Y,J,H,K$, 
\citealt{hill11}; though we note this is not necessary for SFM).


Despite the fact that the SDSS and UKIDSS fluxes were not obtained using 
the same types of apertures, the SFM method could still be used to calculate photo-z. 
However, an improvement can be made by scaling the SDSS fluxes to better 
match the UKIDSS fluxes prior to using SFM.
This allows for a greater number of potential low $\chi^2$ matches.
For each source, the $i$-band flux within a $1.4''$ circular aperture was estimated using 
both the de-Vaucouleurs and exponential profile fits 
(\S~4.4.5.5 of \citealt{stoughton02}) and combined with 
weights \textsc{fracdev} and $1-$\textsc{fracdev}, respectively 
(analogous to \textsc{cmodel}, \citealt{sdssDR2}). 
This is effectively a (PSF-deconvolved) circular-aperture flux estimated 
using a galaxy profile model. 
The ratio between this flux and the $i$-band \textsc{modelmag}
was determined, and used to scale the SDSS fluxes, for each source. 
\newtext{After this process, both the SDSS and UKIDSS fluxes represent 
  similar size apertures. It is not ideal aperture-matched photometry 
  but it is sufficient for SFM because it is an empirical matching technique.}


The choice of flux errors is important as it sets the relative weights between
the best fitting matching-set galaxies. By guidance from nominal limiting 
magnitudes and by trial and error, the linear errors, in place of Poisson errors, were 
taken to be $(0.84,0.54,0.72,1.08,4.44,3.0,3.0,3.0,3.0)$ nanomaggies for
$(u,g,r,i,z,Y,J,H,K)$. 
The fractional errors ($a_k$) were taken to be 0.05 in the $u$-band
and 0.02 in all the other bands. 
The $u$-band was down weighted by this because of the 
signicantly larger spread in these magnitudes compared to a median 
SDSS magnitude for each source. This spread primarily reflects variations in star-formation
rate (SFR) with less constraint on photo-z. 

For any missing measurements (mostly where no UKIDSS match was available), 
the error is set to a extremely large value and the flux to zero 
(i.e.\ to give no effect on normalization or chi-squared). 
In addition, for each source the magnitudes (SDSS and UKIDSS,
separately) were compared to a median value, and any measurements 
outside a large tolerance were set as if missing. 
Thus, some bad measurements are accounted for and do not affect
the photo-z determination. 

The matching set was chosen as a subset of the sample with 
all the following criteria: \textsc{zwarning}$ = 0$, 
spectroscopic classification as a galaxy,  
valid photometric measurements in all nine bands, 
and $0.002 < \zeta < 0.5$. 
The matching-set galaxies were binned in $\zeta$ with 
binsize of 0.002. The weights $W_j$ for all ($n_{\rm bin}$) 
galaxies in a bin were set to $100/n_{\rm bin}^{\alpha}$ with
$\alpha = 0.5$ with a maximum weight of 20 (when $n_{\rm bin} \le 25$). 
The choice of $\alpha = 0.5$ is a compromise between 
equal weight to all galaxies ($\alpha = 0$) and effectively setting 
equal weight to all bins ($\alpha = 1$). 
The number of matching-set galaxies was 194\,183. 


\subsection{Selecting reliable galaxy redshifts}
\label{sec:reliable-photo-z}

\begin{figure}
\includegraphics[width=\singlecolsize\textwidth]{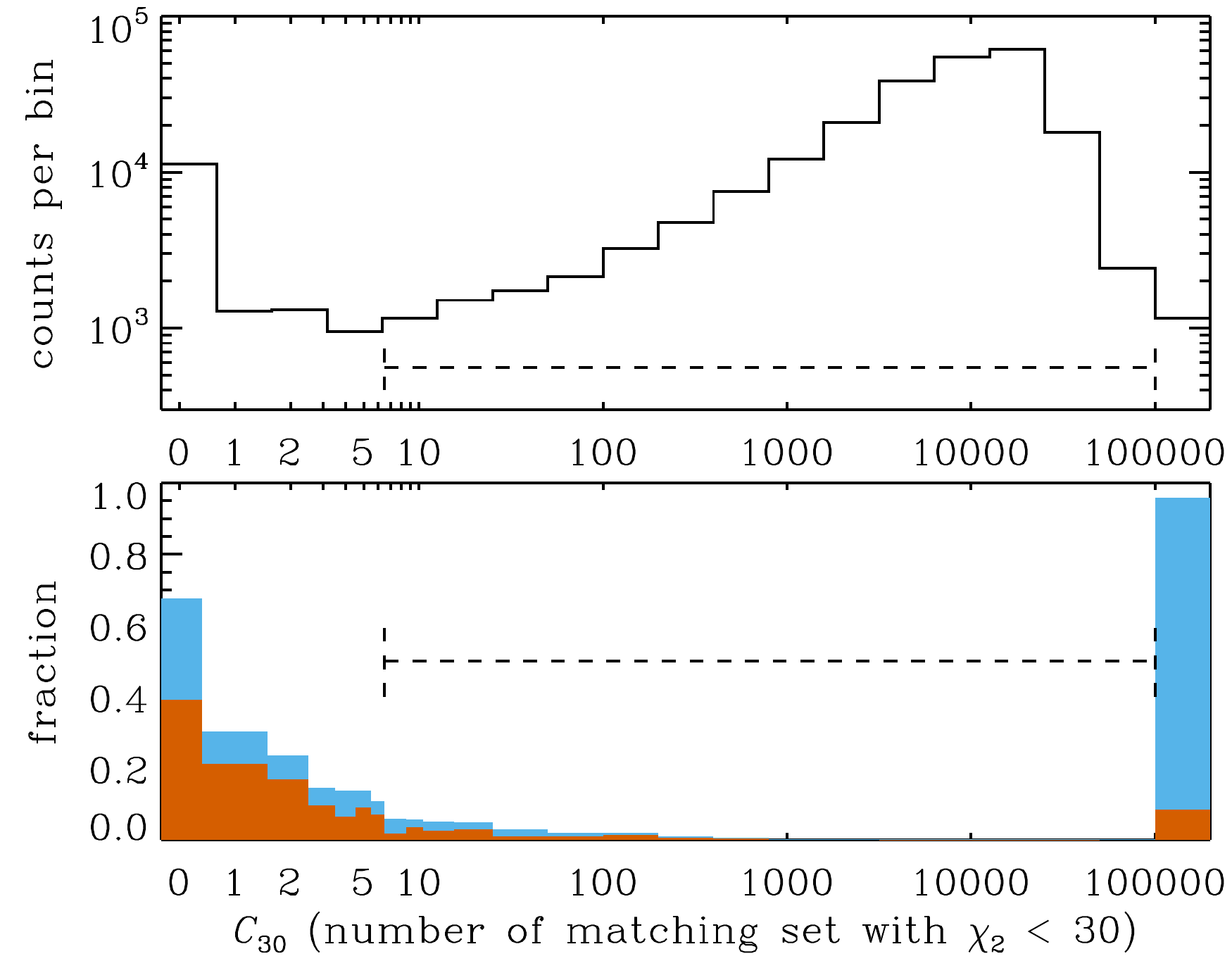}
\caption{Upper panel: histogram in $C_{30}$ of the photo-z measurements.
  Lower panel: fraction of sources (stacked bar chart) classified as stars (orange) and fraction with 
  a redshift offset between the spec-z and photo-z given by $|\Delta \ln (1+z)| > 0.08$ (blue).
  The range used to select reliable photo-z is shown by the dashed line in both panels.}
\label{fig:count-chisq}
\end{figure}

In order to assess the reliability of the photo-z, 
the number of matches with $\chi^2 < 30$ is considered, 
hereafter $C_{30}$, along with the nominal error 
($\zeta_{i,\mathrm{err}}$, Eq.~\ref{eqn:nominal-error}). 
Figure~\ref{fig:count-chisq} shows the distribution in 
$C_{30}$ in the upper panel. Taking only photo-z measurements 
where there is also a reliable spec-z, 
the lower panel indicates the fraction of sources 
spectroscopically classified as stars and the fraction where there is a significant 
redshift offset $|\Delta \ln (1+z)| > 0.08$ between photo-z and spec-z. 
To eliminate poor photo-z and stars from the sample, 
only sources with $6 < C_{30} < 100000$ were set as having reliable photo-z. 

\begin{figure}
\includegraphics[width=\singlecolsize\textwidth]{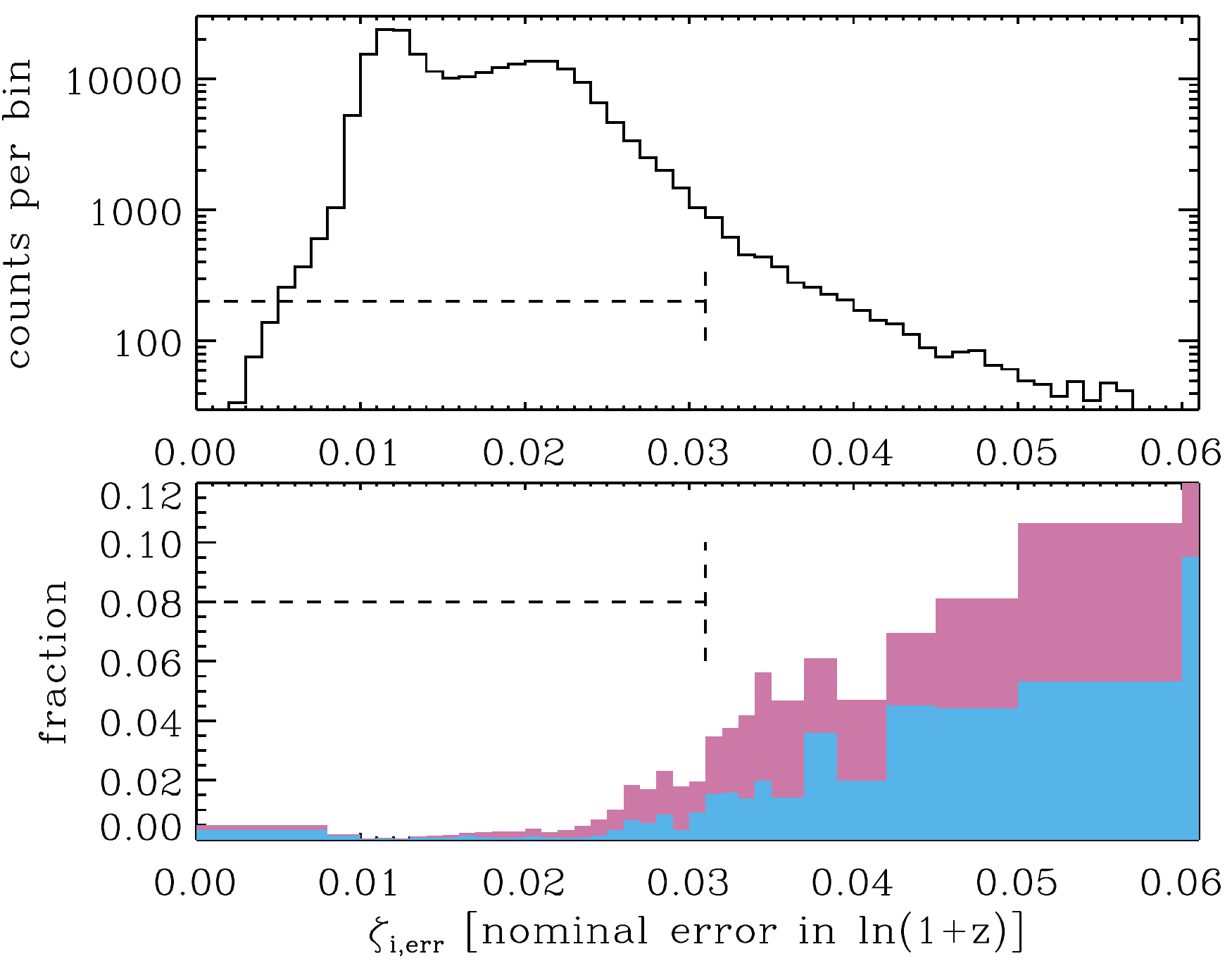}
\caption{Upper panel: histogram in $\zeta_{i,\mathrm{err}}$ of the photo-z measurements.
  Lower panel: fraction of sources (stacked bar chart) with 
  a redshift offset between the spec-z and photo-z given by $|\Delta \ln (1+z)| > 0.08$ (blue),
  and fraction with offset between 0.06 and 0.08 (pink).
  The range used to select reliable photo-z is shown by the dashed line in both panels.
  The sample used for this figure was selected to have $6 < C_{30} < 100000$, i.e., 
  within the reliable range shown in Fig.~\ref{fig:count-chisq}.}
\label{fig:pz-err}
\end{figure}

Figure~\ref{fig:pz-err} shows the distribution in 
$\zeta_{i,\mathrm{err}}$. As expected, the fraction of photo-z outliers 
generally increases with the nominal error estimate. 
To select sources with reliable photo-z, we additionally 
applied the criteria $\zeta_{i,\mathrm{err}} < 0.031$. 
In total, 93.1\% of the 245\,645 sources from Section~\ref{sec:sample-selection} 
have reliable photo-z; and 98.3\% of spec-z confirmed galaxies. 
Thus most of the 6.9\% of the sample without reliable photo-z
are expected to be stars. Of the reliable photo-z sample, only 1 in 1000
are spec-z confirmed stars; and of those with photo-z $>0.04$, only 1 in 3000. 
Thus without using any photometric profile information, 
these have a very low stellar contamination.

\subsection{Redshift accuracy}

\begin{figure}
\centerline{\includegraphics[width=\singlecolsize\textwidth]{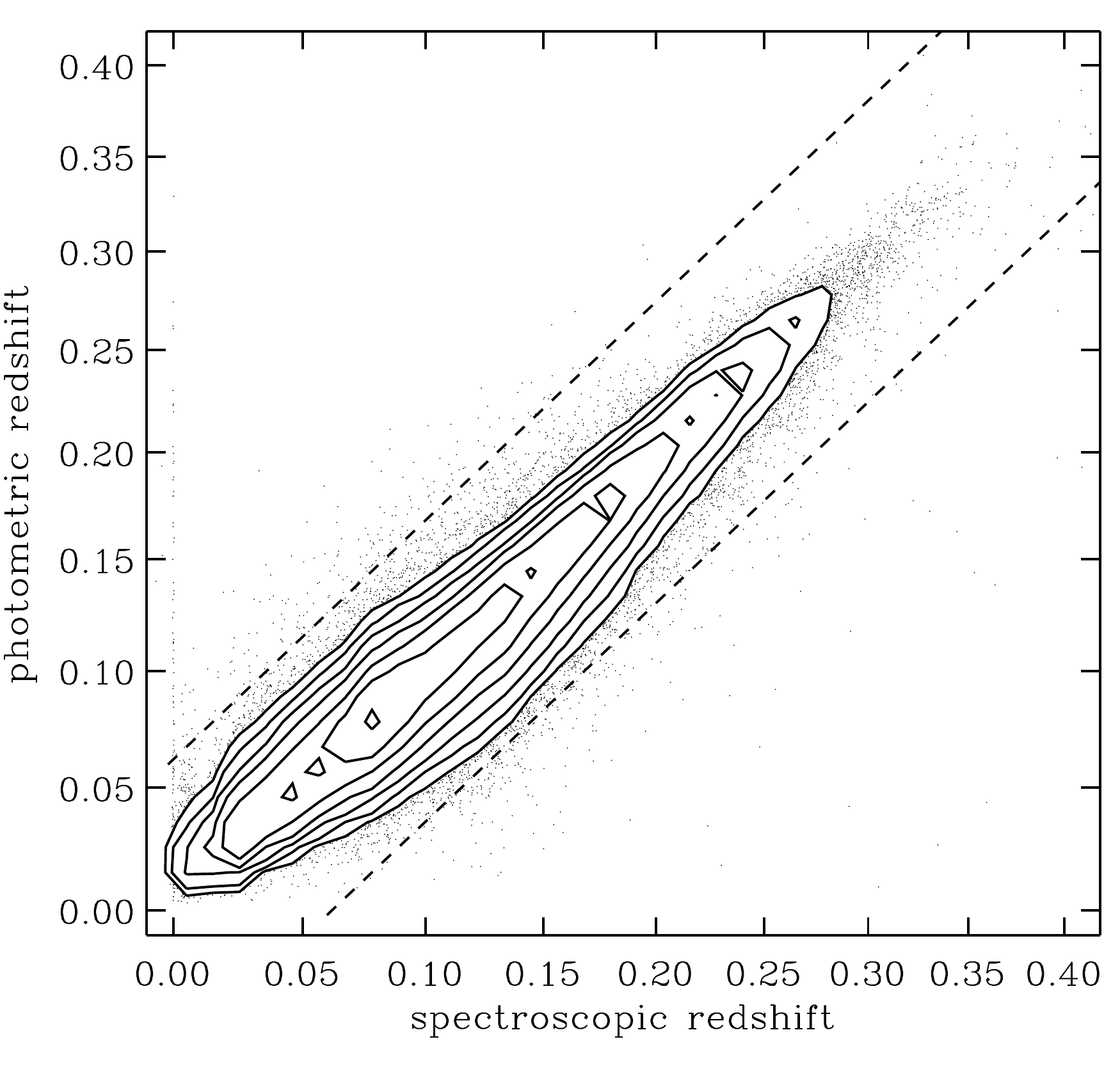}}
\caption{Distribution of photo-z versus spec-z for sources with reliable
 photo-z and spectroscopic \textsc{zwarning}$= 0$. The solid-line contours show
  logarithmically spaced densities with a factor of two between each contour. 
  98\% of the data are represented by the contours, and 2\% by the points.
  The dashed lines show the limits of $|\Delta \ln (1+z)|=0.06$ (99.7\% of the data).} 
\label{fig:spec-z-photo-z}
\end{figure}

Figure~\ref{fig:spec-z-photo-z} shows a plot of spec-z versus photo-z for 
the reliable photo-z sample. The correlation is good with
no obvious bias (considering that the contours represent 
logarithmically-spaced number densities). The 68th percentile ($\sigma_{68}$) of 
$|\Delta \ln(1+z)|$ is 0.0125, and 
the 95th percentile ($\sigma_{95}$) is 0.030. 

These SFM ($ugrizYJHK$) redshifts were compared 
with the empirical photo-z for SDSS 
from \citet{beck16}. The latter were determined using 
`local linear regression' for redshift 
as a function of the $r$-band magnitude and 
four colours ($u-g$, $g-r$, $r-i$, $i-z$). 
For the Beck et al.\ photo-z to spec-z difference, 
$\sigma_{68} = 0.0144$ and $\sigma_{95} = 0.0357$
using a comparison sample of photo-z for which 
the Beck et al.\ analysis defines as reliable 
(\textsc{photoerrorclass}$ = 1$). 
Thus, the SFM photo-z have a 25-30\% reduction 
in the error variance compared to the Beck et al.\ photo-z. 
This is because of the addition of the near-IR data 
rather than the method, which is a similar empirical 
method using a $\chi^2$ matching formalism. 
In any case, the Beck et al.\ photo-z were not determined
for SDSS sources photometrically classified as stars.
Therefore we only use the SFM photo-z that also do not
have a magnitude prior.

\subsection{Final galaxy sample selection}

For the 245\,645 sources from \S~\ref{sec:sample-selection},
redshifts are assigned with the following priority: 
reliable SDSS spec-z, 2MRS, and then reliable photo-z (\S~\ref{sec:reliable-photo-z}),
with 86.1\%, 0.2\%, and 9.4\%, respectively. 
The remainder are assumed to be stars or quasars. 
Redshifts are converted from the heliocentric to the CMB frame,
used hereafter. 
The galaxy sample selection for this study is then 
$0.04 < z < 0.15$. 

\begin{figure}
\includegraphics[width=\singlecolsize\textwidth]{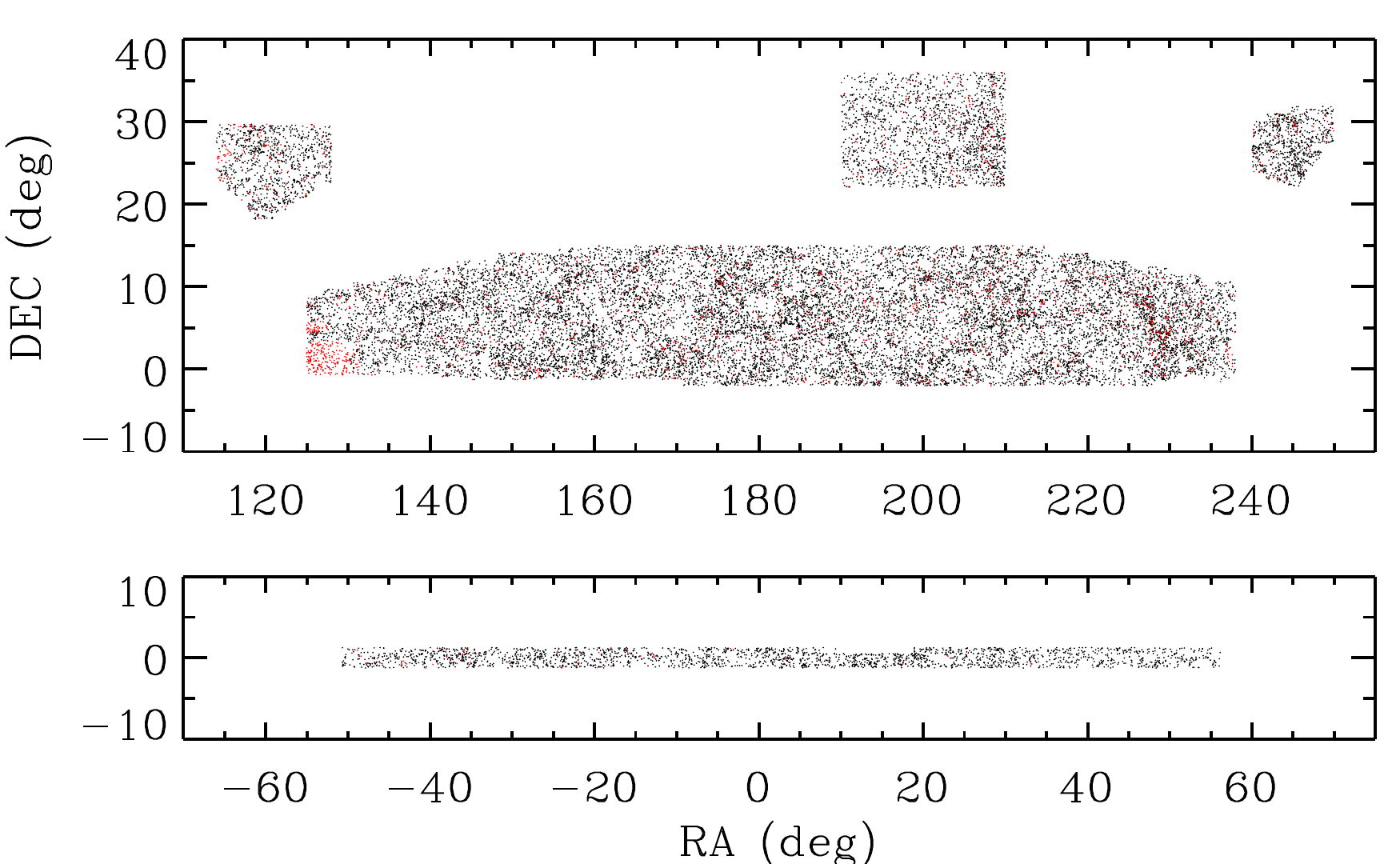}
\caption{Sky positions of the galaxy sample with $17.3 < r < 17.5$ 
  demonstrating the $ugrizYJHK$ coverage ($2300\,\sqdeg$) from SDSS and UKIDSS.
  The black dots represent galaxies with spec-z and the red dots with
  photo-z.}
\label{fig:regions}
\end{figure}

Compact galaxy candidates were visually inspected and 
27 sources whose photometry was significantly 
affected by scattered light (missed by the masking procedure)
or a bright neighbour were rejected.  
In addition, 138 sources with GAIA photometry that had 
$\mathrm{S/N} > 5$ for their parallax measurement or 
$\mathrm{S/N} > 10$ for their proper motion were rejected. 
The latter cut was set high because there were many obviously extended 
sources that had S/N between 5 and 10, some of which may be 
star-galaxy blends but in other cases it was not clear. 
After the redshift limits and these cuts, 
a sample of 163\,186 sources remained.
Figure~\ref{fig:regions} shows the sky positions of a subset of
the sample. 

\section{Results and discussion}
\label{sec:size-mass}

\subsection{Stellar masses}

Stellar masses ($M_{*}$) were determined for the spec-z and photo-z sample using the 
the method of \citet{sedgwick19} (updated from \citealt{bryant15}) such that 
\begin{equation}\label{eq:mass}
  \log M_{*} = -0.4 i + 0.4 {\cal D} + f(\mathrm{z}) + g(\mathrm{z}) (g-i)_{\rm obs}
\end{equation}
where $i$ is the $i$-band \textsc{cmodel} magnitude, ${\cal D}$ is the distance modulus,
and $(g-i)_{\rm obs}$ is the observed colour using Petrosian magnitudes.
This effectively folds the mass-to-light ratio and k-correction into the same formula.
\newtext{These magnitude types were chosen as \textsc{cmodel} represents the best estimate
  of the total flux derived from profile fits, while Petrosian magnitudes provide
  the best unbiased estimate of galaxy colours.}
The functions $f(\mathrm{z})$ and $g(\mathrm{z})$ are calibrated to a stellar-population
fitted set of stellar masses.
In this case, the functions were calibrated using the GAMA survey photometry and products
\citep{baldry18} with the stellar masses from the \textsc{magphys} code \citep{dCCE08}
applied by \citet{driver16}. 
These masses were used in order to be consistent with the high-z comparison sample. 

The functions used in this paper are given by
$f(z) = 1.008  - 3.531 z + 21.64 z^2 - 35.28 z^3$ and
$g(z) = 0.8132 + 3.304 z - 30.26 z^2 + 55.60 z^3$.
The calibration is valid over the range $0 < z <0.35$.
The stellar initial mass function (IMF) assumed was the \citet{Chabrier03} IMF. 

We note that dynamical modelling using resolved spectroscopy \citep{cappellari12,posacki15} and
detailed stellar population analysis \citep{CvD12,labarbera13} 
have suggested that there is a bottom-heavy IMF in early-type galaxies that have high velocity dispersion.
This is relevant for massive and/or compact galaxies.
\newtext{For example, \citet{ferre-mateu17} modelled the spectra of two compact galaxies using steep IMF slopes.}
However, \citet{collier18} showed that the mass obtained using gravitational lensing for four low-redshift ($z < 0.07$) 
massive ellipticals was consistent with a \citet{Kroupa01} IMF. Strong lensing of these galaxies 
is dominated by their stellar mass \citep{SmithLucey13} and therefore is an accurate constraint on 
stellar mass-to-light ratios. 
\newtext{Regardless of the variation or not of the IMF between galaxies, it is appropriate to assume a universal
  IMF for distribution functions or relationships involving stellar mass because the extra mass, if any, for some galaxies
  has little impact on broadband photometry and, for transparency and simplicity, the stellar mass can be considered as a `modified luminosity'.
  This is particularly the case here because we only use $g$- and $i$-bands to estimate the stellar mass. 
  Stellar masses obtained assuming a Kroupa IMF (Salpeter IMF) are about 0.05\,dex (0.25\,dex) higher than using the Chabrier IMF
  \citep{collier18}.}


The stellar mass approximation, above, uses observed colour.
To define red galaxies, we use 
\begin{equation}
  (u - r)_{\rm adj} = u - r - 4.5 \ln(1+z) \: .
\end{equation} 
This approximates the $k$-and-evolution correction for dividing the red and blue sequences
of galaxies at a rest-frame $u-r \sim 2$.

Specific star formation rates (SSFRs) were obtained for the sample
from the SDSS table \textsc{galSpecExtra}. These were derived from spectral emission
lines \citep{brinchmann04} with aperture corrections using photometry \citep{salim07}.
For sources without a match, e.g.\ with photo-z only, the nearest SFM matching-set galaxy
was used for the SSFR. This provided SSFR measurements for 98.6\% of galaxies 
with $\log M_{*} > 10$.
Red galaxies were defined using $(u - r)_{\rm adj} > 2.0$ while quenched galaxies
were defined using $\log (\mathrm{SSFR/yr}^{-1}) < -11.2$ (or using the red galaxy definition for the 1.4\% without
SSFR measurements) for this $z<0.15$ sample.


\subsection{Half-light radii}
\label{sec:half-light}

Various half-light radii ($r_{50}$) are provided for SDSS.
The SDSS pipeline computes Petrosian $r_{50}$ (circular), de Vaucolueurs and exponential
profile fits (elliptical) \citep{stoughton02}.
Only the profile fits take account of the PSF.
\newtext{The SDSS pipeline ``also takes the best-fit exponential and de Vaucouleurs fits
  in each band and asks for the linear combination of the two that best fits the image''
  \citep{sdssDR2}. This provides a coefficient (clipped between zero and one), called
  \textsc{fracdev} ($f_{\rm dev}$), that gives the relative flux contribution of the de Vaucouleurs fit,
  with $1-f_{\rm dev}$ giving the contribution of the exponential fit. 
  This `composite model' is effectively a non-simultaneous fit of the two profile functions 
  because the profiles are fitted separately before being combined.}
In addition, \citet{simard11} provides bulge plus disk simultaneous fits, and Sersic fits for 
SDSS galaxies with spec-z, using \textsc{gim2d} \citep{simard02}.
These are provided for the $g$- and $r$-bands.

\begin{figure}
\includegraphics[width=\singlecolsize\textwidth]{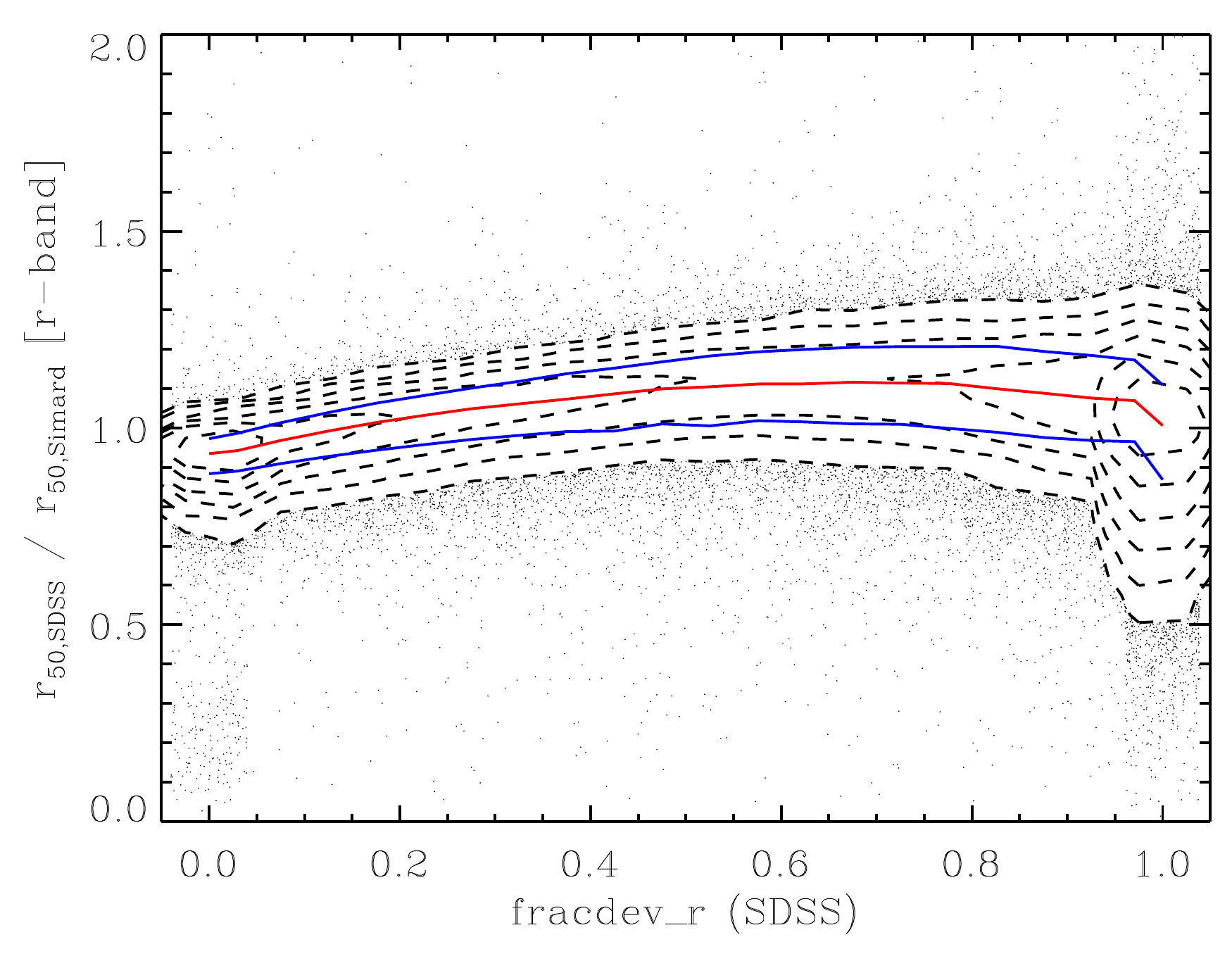}
\caption{Comparison between the half-light radii determined using
  the SDSS profile fits (weighted average) with the radii from \citet{simard11}.
  The dashed contours and points show the bivariate distribution, of the galaxy sample,
  in the ratio between the radii measurements versus \textsc{fracdev\_r},
  which is determined by the SDSS pipeline. (Note jitter has been added to
  \textsc{fracdev\_r} to spread out the data at 0 and 1.) 
  The red and blue solid lines show the median, 16th and 84th percentiles of the ratios 
  binned in \textsc{fracdev\_r}. For the whole sample, 93\% of the galaxies have measurements
  differing by less than 0.1\,dex (ratios between 0.79 and 1.26).}
\label{fig:size-comparison}
\end{figure}

Since the \citeauthor{simard11} fits are not provided for sources without spec-z, or for the $i$-band,
we used the half-light radii from SDSS with a weighted geometric average of the
two models for each band:
\begin{equation}
  \log \reff = f_{\rm dev} \log r_{\rm dev} + (1 - f_{\rm dev}) \log r_{\rm exp} \mbox{~~~.}
  \label{eqn:half-light}
\end{equation}
Figure~\ref{fig:size-comparison} shows a comparison between the the radii from these non-simultaneous fits
with the simultaneous fits (de Vaucolueurs bulge plus exponential disk) of \citeauthor{simard11}.
There is generally good agreement with 93\% of galaxies within 0.1\,dex, and
98.5\% within 0.2\,dex.
The geometric weighted mean for SDSS agreed marginally better with the comparison sample
than the linear weighted mean.

The final $\reff$ we used was taken as the mean between the $r$ and $i$-band.
This was to increase the fidelity, ensuring that compact galaxies had to have small fitted 
radii in two bands, while avoiding the bluer $g$-band that crosses the 4000\AA\ break 
at $z>0$ and the $z$-band which is of lower S/N.
The mean rest-frame effective wavelength of the $r$- and $i$-bands is $\sim$6000--6500\AA\ for
$0.04 < z < 0.15$.
For a few sources, half-light radii were clipped to a minimum of $0.3''$. 
In total, there were 34 unresolved sources with mean radius $<0.35''$ and that were 
photometrically classified as a star by the SDSS pipeline.
Of the 18 that had estimated stellar masses $\log M_{*} > 10$, all but one 
had log SSFR $> -10$. This sample is of interest as candidate compact starbursts 
but has no significance for the number densities of the compact quiescent population. 
Note \citet{taylor10} used a minimum of $0.75''$ for the $z$-band sizes,
however, we find that reasonable agreement between the \citet{simard11} 
and pipeline half-light radii, and between $i$- and $r$-bands, extends to lower values.
This is testament to the accuracy with which the PSF is measured and accounted for
in the galaxy profile fitting. \newtext{See Appendix for further comparisons.}

\subsection{Size-mass distribution}
\label{sec:size-mass-distribution}

\begin{figure}
\includegraphics[width=\singlecolsize\textwidth]{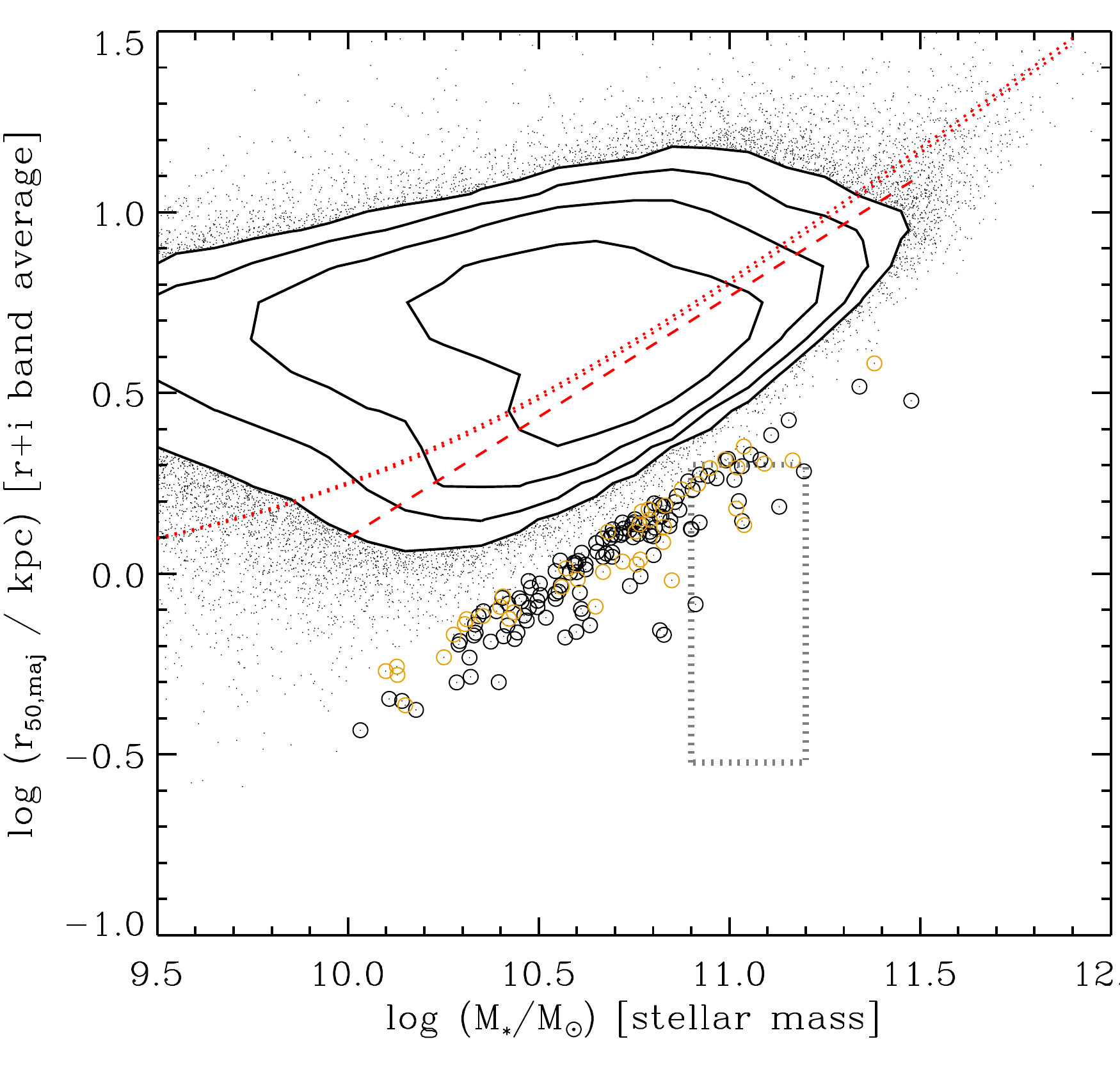}
\caption{Size-mass distribution for the galaxy sample ($0.04 < z < 0.15$).
  The contours and points represent the majority of the galaxy sample,
  while the circles represent galaxies with $\log \Sigma_{1.5} > 10.5$
  (black with spec-z and orange with photo-z).
  The dashed line shows $\log \Sigma_{1.5} = 9.85$,
  which is the median for the quiescent population at $\log (M_{*} / \Msun) > 10$.
  The red dotted lines show the two-power-law fits from \citet{lange15} for the $r$- and $i$-bands $\reff$ 
  with the early-type division using Sersic index and $u-r$.
  \newtext{The grey dotted box shows the ultra-compact massive galaxy (UCMG) selection with a $\reff<2.0$ cut.}}
\label{fig:size-mass}
\end{figure}

Figure~\ref{fig:size-mass} shows the size-mass distribution of the SDSS-UKIDSS selected
sample. 
Following \citet{barro13}, we define a projection in the size-mass distribution that runs
perpendicular to the quiescent size-mass relation at high masses.
This is given by
\[
 \log \Sigma_{1.5} = \log (M_{*} / \Msun) - 1.5 \log (\reff / \mathrm{kpc}) \mbox{~~~.}
 \label{eq:sigma-1pt5}
\] 
Thus $\Sigma_{1.5}$ has units of $\Msun\mathrm{kpc}^{-1.5}$ and uses physical half-light radius. 
Compact galaxies are shown with circles in the figure that have $\log \Sigma_{1.5} > 10.5$ 
\newtext{(sample images are shown in Fig.~\ref{fig:images1}).}
This represents only 0.12\% of the sample at $\log M_{*} > 10$.
Note we prefer to use the major axis to define radii and $\Sigma_{1.5}$ because 
cicularization of effective radii is strongly inclination dependent \citep{DevourBell17}.

\begin{figure}
\includegraphics[width=\singlecolsize\textwidth]{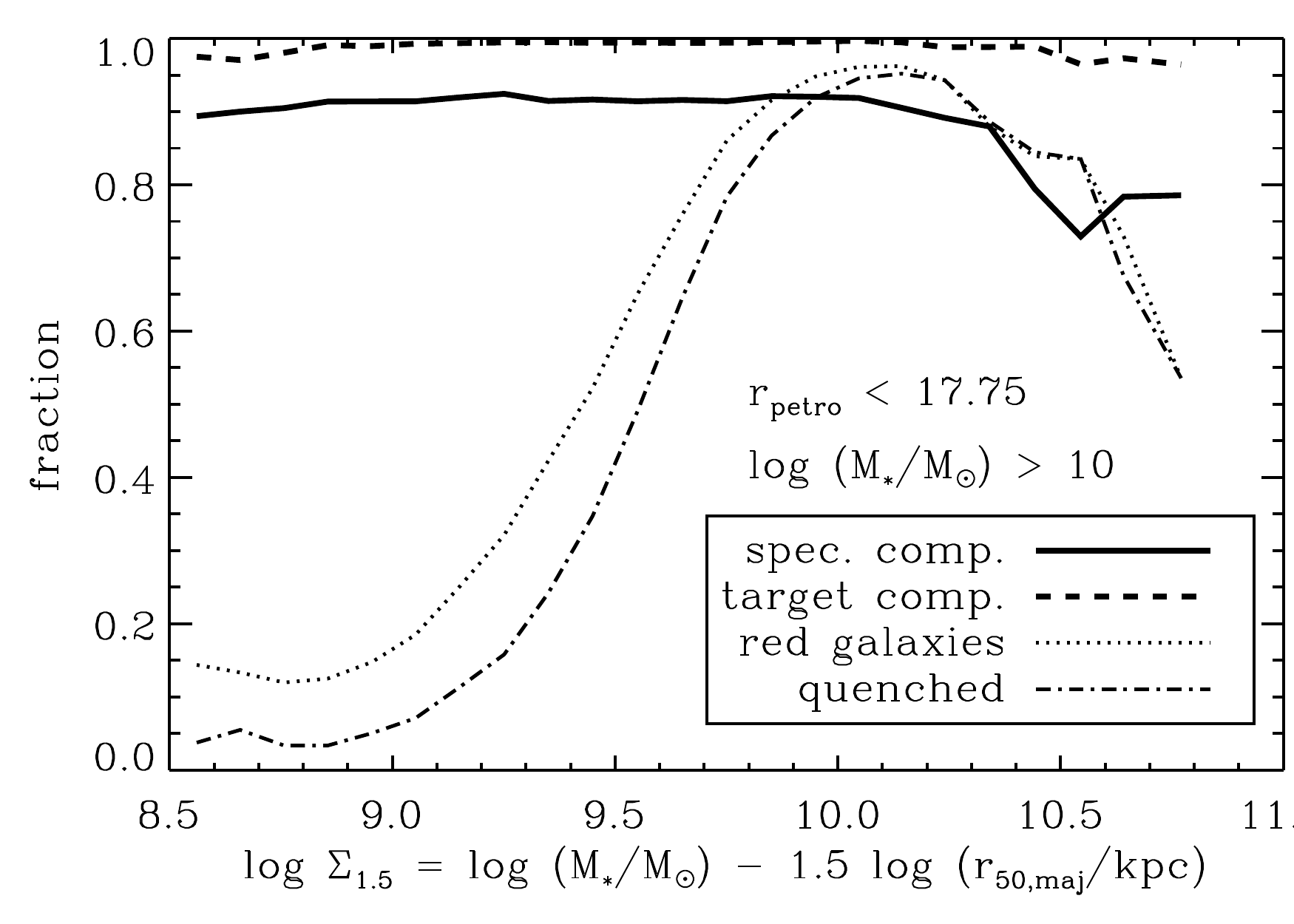}
\caption{Completeness as a function of $\log \Sigma_{1.5}$.
  The solid line shows the spectrosopic completeness of the sample, while the dashed
  line shows the SDSS target completeness i.e.\ the fraction of galaxies
  that were part of the SDSS main galaxy sample, primarily. Also shown are the fraction
  of red galaxies [$(u - r)_{\rm adj} > 2.0$] and quenched galaxies
  [$\log ({\rm SSFR/yr}^{-1}) < -11.2$].}
\label{fig:completeness}
\end{figure}

The orange circles in Fig.~\ref{fig:size-mass} represent the 26\% of the compact galaxy sample that have photo-z only.
For compact galaxies, there is a significantly larger fraction with photo-z compared to the full sample. 
This demonstrates a bias against compact galaxies in the SDSS spectroscopic sample. 
Figure~\ref{fig:completeness} shows the spectroscopic and target completeness
as a function of $\log \Sigma_{1.5}$
($r < 17.75$ to minimise the issue of photometric scatter across the 17.77 selection boundary
between SDSS data reduction versions).
The target completeness, in this paper, is the fraction of sources that were selected as targets for SDSS
spectroscopy regardless of whether a spectrum was obtained. 
The spectrocopic (target) completeness is 91.6\% (99.4\%) in the normal $\log \Sigma_{1.5}$ range
(8.8--10.3). This drops to 75.8\% (96.7\%) for $\log \Sigma_{1.5} > 10.5$.
This means that the majority of the bias against compact galaxies is due to 
`fibre collisions' and not the SDSS photometric selection.
Fibre collisions were caused by the restriction of how close fibres could be placed on the spectroscopic plate
\citep{stoughton02}.
The bias arises because compact galaxies are more likely to be found in
proximity to other SDSS galaxies \citep{trujillo14} (\S~\ref{sec:enviro}). 


The fraction of red galaxies and quenched galaxies as a function of $\log \Sigma_{1.5}$
is also shown in Figure~\ref{fig:completeness}.
The red fraction is higher than the quenched fraction for less compact galaxies.
This is because the estimate of SSFR takes account of internal dust attenuation
whereas the $u-r$ colour does not. In other words,
some dusty star forming galaxies appear as red but not as quenched.
The red/quenched fraction rises from less than 0.2 at $\log \Sigma_{1.5} \sim 9.0$
to 0.95 at $\sim 10.2$. This range is where the majority of galaxies are distributed. 
For compact galaxies ($\log \Sigma_{1.5}>10.5$), the red/quenched fraction drops below 0.85.
This is due to a small population of quasars and starburst galaxies,
for which, the SDSS half-light radii likely do not represent the overall stellar population
within each galaxy. 

\subsection{High-z comparison}
\label{sec:high-z}

In order to compare the distribution functions from the SDSS-UKIDSS sample at $z \sim 0.1$
to high redshift, we used the 3D-HST survey data. 
These are five fields with near-IR grism spectra \citep{brammer12} and photometry from HST
primarily from the CANDELS treasury programme \citep{grogin11}.
The construction of the photometric catalogues, including space and ground-based imaging,
is described in \citet{skelton14}. The science area (\textsc{use\_phot=1}) of the combined
fields is $0.249\,\sqdeg$. The sources were assigned photo-z, ground-based spec-z or
grism redshifts \citep{momcheva16}. 

Structural parameters using Sersic models were determined for the sources by \citet{vanderwel12} in
the near-IR F160W ($H$) and F125W ($J$) bands. These included half-light radii and
profile-integrated magnitudes. Stellar masses and SSFR were determined by \citet{driver18}
using \textsc{magphys}. For this paper, the stellar masses were corrected for the difference
between the F160W magnitude \citep{skelton14} used in the \textsc{magphys} fitting and
the Sersic F160W magnitude. F125W was used if F160W was not available ($< 0.5$\% of sources). The same
procedure was used in \citet{vanderwel14}, except here we use the \textsc{magphys} stellar masses.
Selecting galaxies with $0.5 < z < 2.5$, $\log M_{*} > 10$, \textsc{use\_phot}$\,= 1$,
and rejecting poor fits visually checked by \citet{vanderwel14}, produced a
volume-limited sample of 9757 galaxies (over $4.8 \times 10^6 \, \mathrm{cMpc}^3$).

The half-light radii were obtained using the F160W band ($>99.5$\% of sources),
which has a pivot wavelength of 15400\AA. 
This matches the rest-frame wavelength used for the low-redshift sample at $z \sim 1.5$. 
At higher redshifts, the rest-frame wavelength drops below 6000\AA\ and thus the measured radii are  
potentially affected more significantly by younger stellar populations. 
This is mitigated by the fact that the colour gradients in galaxies are close to zero
at $z \sim 2$--2.5 \citep{suess19}.
Therefore, measurements in F160W in the high-z sample can be reasonably compared to the
low-z measurements without significant concern regarding the dependence of radii on rest-frame
wavelength.

Star-forming galaxies form a `main sequence' in SSFR versus stellar mass that evolves 
with redshift \citep{noeske07,salim07,pearson18,popesso19}. 
Therefore, we define `quenched' using a dividing line that evolves with redshift
such that 
\begin{equation}
  \log ({\rm SSFR}_{\rm divide}/{\rm yr}^{-1}) \: = \: -11.2 + 1.2 \ln (1+z) \: ,
  \label{eq:ssfr-cut}
\end{equation}
which corresponds to $-10.4$ at $z=1$ and $-9.7$ at $z=2.5$
[$-11.2 + 2.76 \log(1+z)$]. 
This dividing line is about 1\,dex below the SSFR versus redshift of the main 
sequence for $\log M_{*} > 10$ galaxies. 

\subsection{Distribution functions}

The total volume covered by the SDSS-UKIDSS sample is $54.6\times 10^6 \, \mathrm{cMpc}^3$. 
This is only a volume-limited sample for $\log M_* \ga 10.9$. 
To determine distribution functions including galaxies less massive than this,
1/Vmax weighting was used \citep{Eales93}.
\newtext{Distribution functions were then obtained by summing 1/Vmax in bins.
For the high-z comparison sample, Vmax was the same for all galaxies in a given redshift range.}

\newtext{For the SDSS-UKIDSS sample, the maximum luminosity distance ($D_L$) that a galaxy, of the same type and luminosity, 
  could be seen at and remain within the flux limit is given by 
  \begin{equation}
     5 \log(D_{L,\mathrm{max}}) =  5 \log(D_{L,\mathrm{obs}}) + 17.8 - r_{\rm Petro} + dK
  \end{equation}
  where $dK$ accounts for the difference in the $k$-correction between the observed and maximum redshifts.
  For the purposes of this paper, sufficient accuracy was obtained with $dK$ set to zero. 
  Then, from $D_{L,\mathrm{max}}$, the maximum redshift ($z_{\mathrm{max}}$) was determined for each galaxy. 
  These $z_{\mathrm{max}}$ values were clipped to lie between a lower limit as a function of stellar mass, 
  that was determined using the 1st percentiles in mass bins, and $z=0.15$, the redshift limit of the sample. 
  Vmax is then given by the comoving volume between 0.04 and $z_{\mathrm{max}}$. 
}
At $\log M_* = 10.0$, Vmax is about $5 \times 10^6 \, \mathrm{cMpc}^3$ for the
least luminous galaxies ($z_{\mathrm{max}} \approx 0.07$), which are quenched galaxies.
Thus the volume covered is larger than, or comparable to, the high-z comparison sample
for all types of galaxies with $\log M_* > 10$. 

\begin{figure}
\includegraphics[width=\singlecolsize\textwidth]{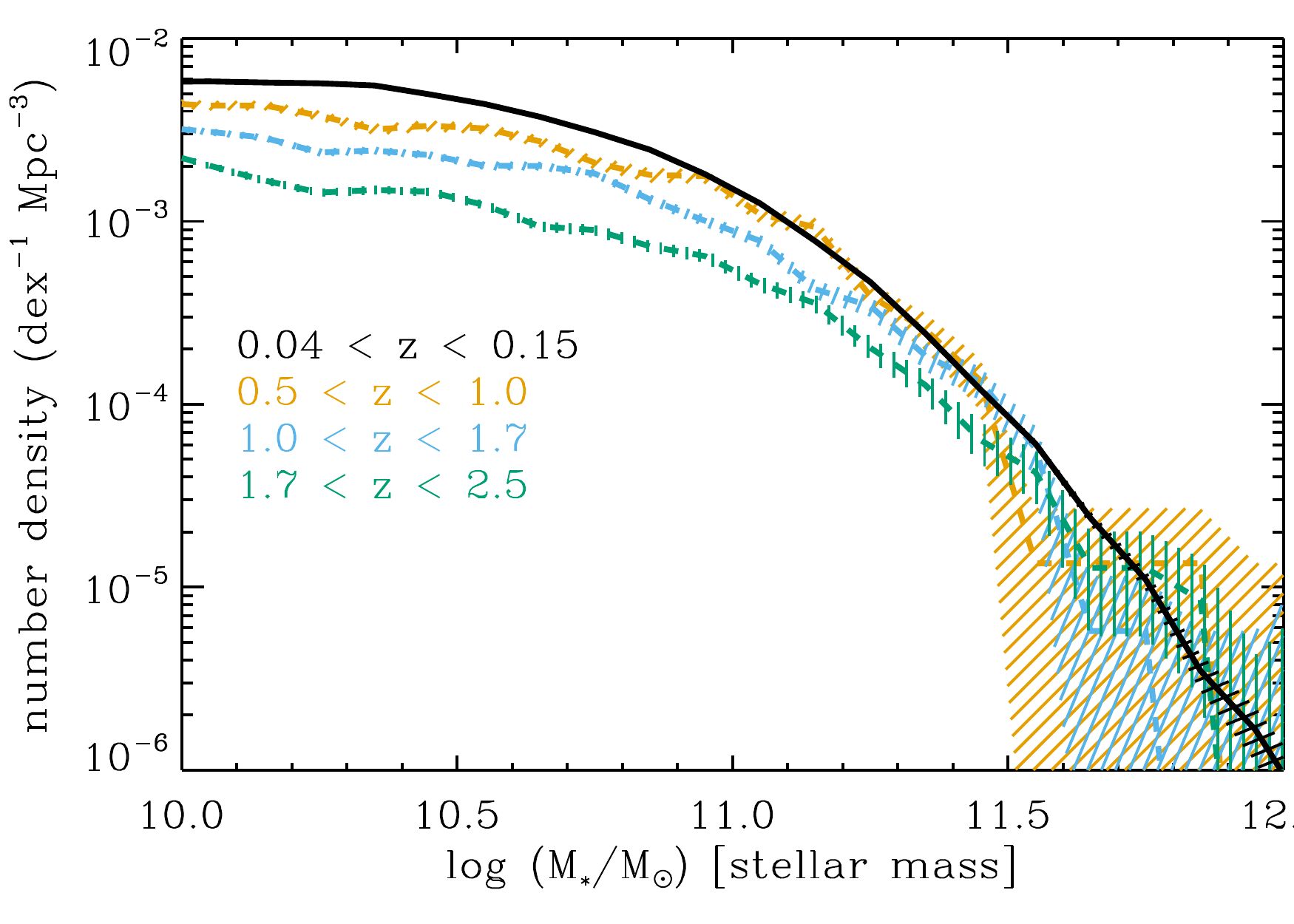}
\caption{Galaxy stellar mass functions.
  The hatched regions represent the Poission counting uncertainties ($\pm\sigma$).}
\label{fig:gsmf-evolve}
\end{figure}

Figure~\ref{fig:gsmf-evolve} shows the galaxy stellar mass functions (GSMFs) for the
SDSS-UKIDSS sample and for the high-z comparison split into three redshift bins.
This demonstrates the trend that most of the highest mass galaxies
($\log M_{*} > 11$) have been in place since $z \sim 2$ with more growth
in number density occurring at lower masses
\citep{davidzon13,muzzin13,mortlock15,huertas-company16,wright18,kawinwanichakij20}.
\citeauthor{kawinwanichakij20} noted that the near absence of observed evolution
in the number density of $\log M_{*} > 11$ galaxies does not mean no mass growth.
This is because stellar mass loss from late-stage stellar evolution 
could be compensated by mass growth from dry mergers or residual star formation. 
They estimated an upper limit of 0.16\,dex for this mass growth of 
supermassive quiescent galaxies from $z = 1.0$ to 0.4. 


\begin{figure}
\includegraphics[width=\singlecolsize\textwidth]{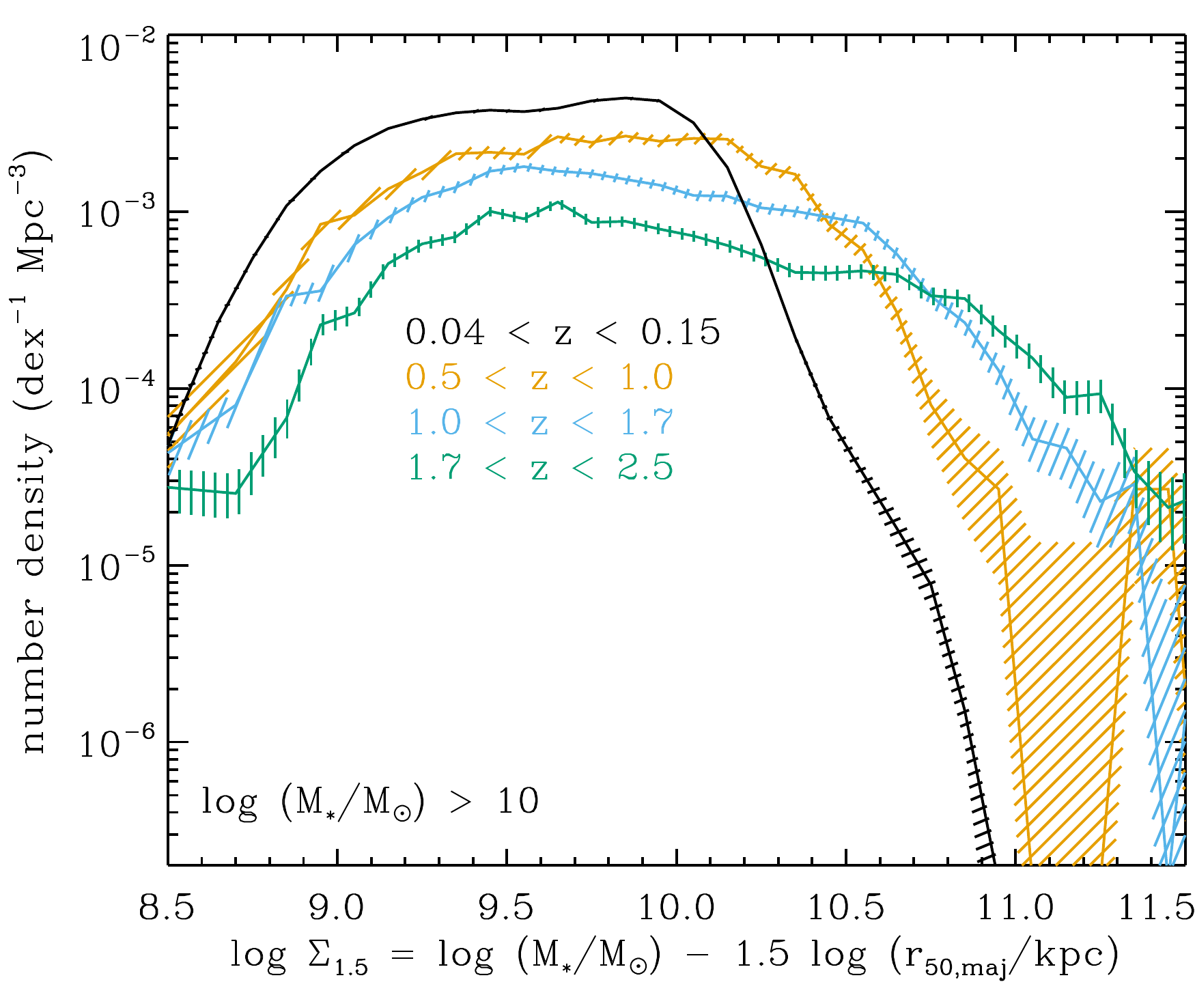}
\includegraphics[width=\singlecolsize\textwidth]{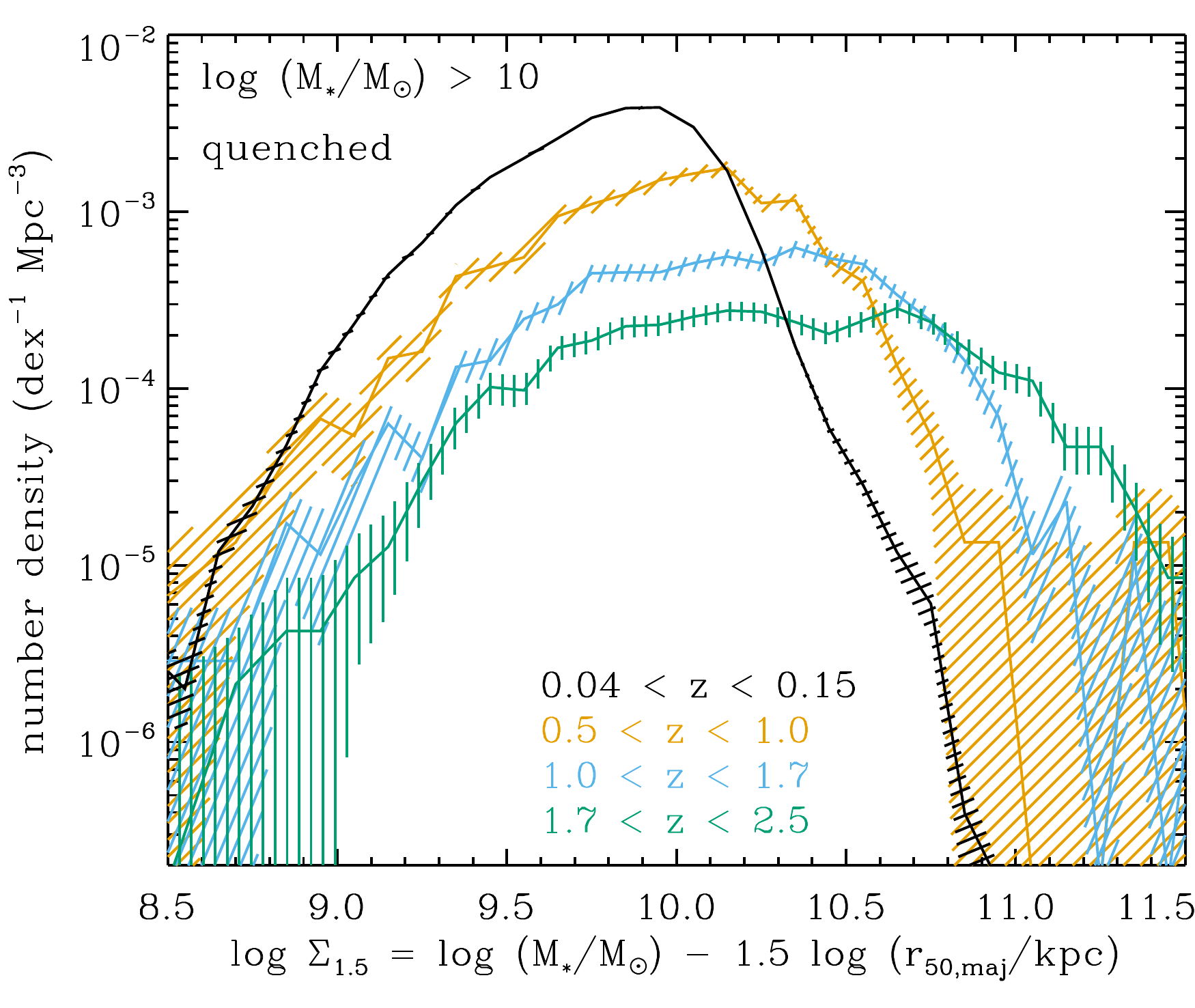}
\includegraphics[width=\singlecolsize\textwidth]{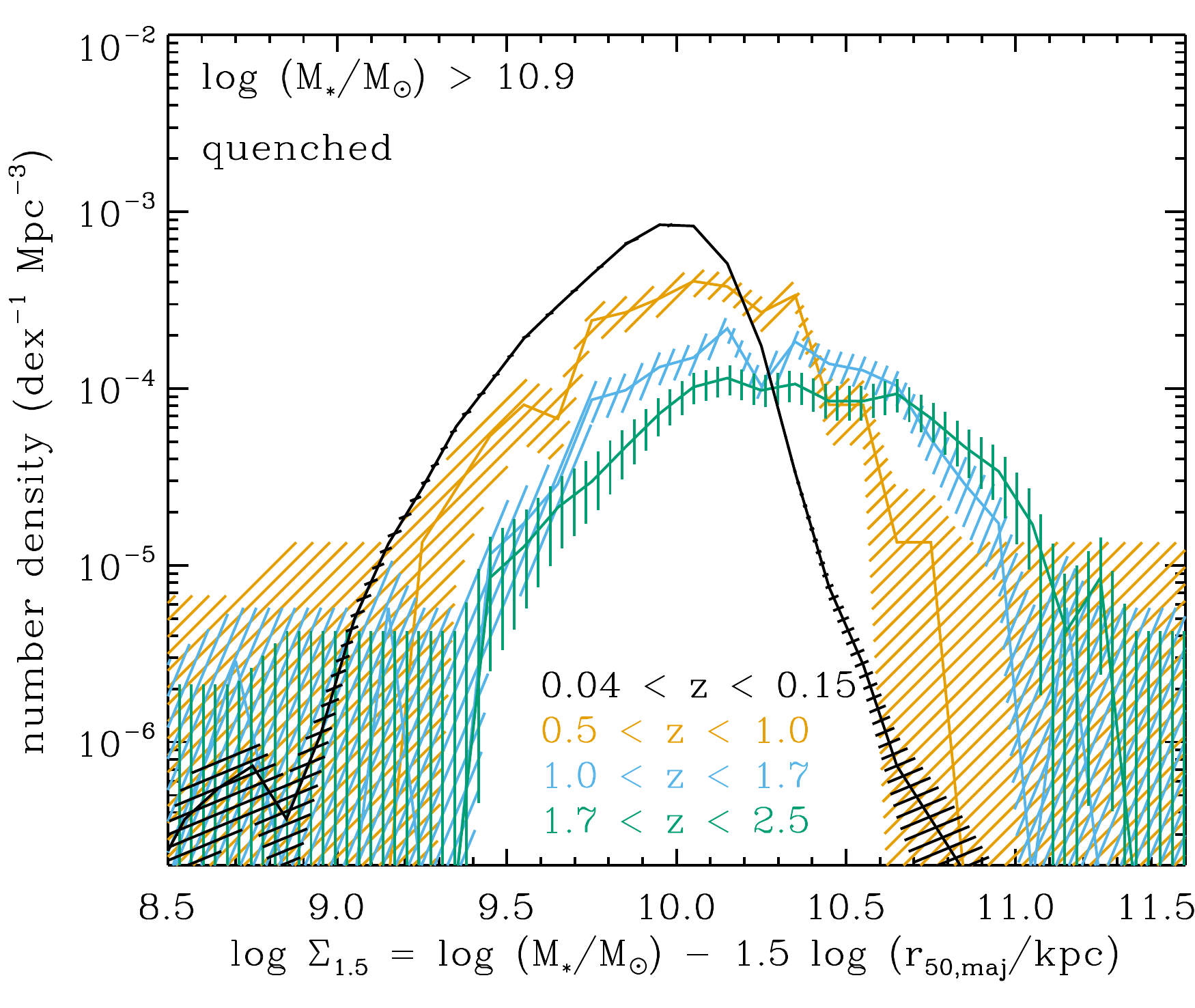}
\caption{$\log \Sigma_{1.5}$ distribution functions.
  The hatched regions represent the Poisson counting uncertainties ($\pm\sigma$).
  Upper panel (a): for all galaxies with $\log (M_{*} / \Msun) > 10.0$.
  Middle panel (b): for quenched galaxies. 
  Lower panel (c): for supermassive quenched galaxies ($\log (M_{*} / \Msun) > 10.9$).}
\label{fig:sigma-func-evolve}
\end{figure}

Size-mass distributions of the high-z sample are shown in figure~5 of \citet{vanderwel14}.
In this paper, in order to demonstrate the evolution of the size-mass distribution in comparison
to the SDSS-UKIDSS sample, we computed the distribution functions in $\log \Sigma_{1.5}$.
In other words, this represents the number density of galaxies running perpendicular
to the high-mass quiescent relation.
Such a representation is suggested by the analysis of \citet{barro13}. 
Figure~\ref{fig:sigma-func-evolve} shows the $\log \Sigma_{1.5}$ distribution functions
for the SDSS-UKIDSS sample and three high-z redshift bins for:
(a) $\log (M_{*} / \Msun) > 10.0$,
(b) quenched galaxies with the same mass limit, and
(c) quenched galaxies with $\log (M_{*} / \Msun) > 10.9$.

The evolution in the $\log \Sigma_{1.5}$ distribution functions
(Fig.~\ref{fig:sigma-func-evolve}) is far more striking than
the GSMF evolution (Fig.~\ref{fig:gsmf-evolve}).
The high $\log \Sigma_{1.5}$ cutoff becomes sharper and moves to lower
values as the galaxy population evolves (from high to low redshift).
At the same time, there is an increase in number density of the less compact galaxies
($\log \Sigma_{1.5} \la 10$).
The pattern is similar for all three selections with narrowing of the functions for
the quenched galaxies, and the supermassive ($\log (M_{*} / \Msun) > 10.9$) quenched galaxies. 

This supports the picture of \citet{barro13} (their figure 6): 
size growth due to minor mergers \citep{naab09,hilz12} for compact quenched galaxies 
that primarily formed at $z \ga 2$, and 
galaxies with normal sizes quenching at all these epochs. 
In this way, the $\log \Sigma_{1.5}$ distribution cutoff at high values becomes steeper. 
In this paper, we have demonstrated this by using a rigorously complete 
low-redshift galaxy selection for compact sources, and 
we also used the major axis for the half-light radii. 
Note that the increase of half-light radii caused by minor merger growth cannot be entirely
due to addition of light at large radii. 
This was demonstrated by \citet{vandokkum14} who showed that the number density of
high-mass-core galaxies, 
with $\log M_* > 10.5$ within the central kpc of each galaxy, decreased from high to low redshift.
This could be explained by a small amount
of stellar-evolution mass loss from the core leading to adiabatic expansion. 

In summary, compact quiescent galaxies can grow primarily through minor mergers 
that increase a galaxy's mass and size.
The size growth is a combination of adding mass to the envelope \citep{huang13} and 
dynamical friction causing the core to expand \citep{naab09}. 
The mass increase can be compensated by mass loss from stellar evolution,
and the core can also grow marginally through adiabatic expansion \citep{vandokkum14}.
It should also be noted that compact galaxies can accrete a disk becoming
early-type spirals \citep{GaoFan20}, 
and then in some cases becoming low-redshift S0 galaxies \citep{graham15,delarosa16,deeley20}.
No doubt there is more than one path to galaxy growth given the stochastic 
nature of hierarchical assembly.
\citet{GaoFan20} estimated, based on a spectral and structural analysis of SDSS galaxies,
that about 15\% of compact quiescent galaxies become 
spirals with a massive bulge with the remaining growing through dry minor mergers
to become massive early types. 

The evolution of the size-mass distribution of galaxies can be examined using cosmological-scale
hydrodynamical simulations. \citet{furlong17} found good agreement for the size-mass evolution
in the EAGLE simulation compared to the \citet{vanderwel14} observational measurements.
In all cases of $z=2$ compact galaxies identified in the simulation,
stars formed at high redshift migrated to larger radii at $z=0$ \citep{furlong17}.
For some of these galaxies, further growth occurred by mergers and star formation.
\citet{wellons15} looked at the formation of supermassive compact galaxies at $z=2$
in the Illustris simulation, finding that they could be formed
by gas-rich mergers at $z < 4$ or by assembly at early times when the universe was much denser.  

\begin{figure}
\includegraphics[width=\singlecolsize\textwidth]{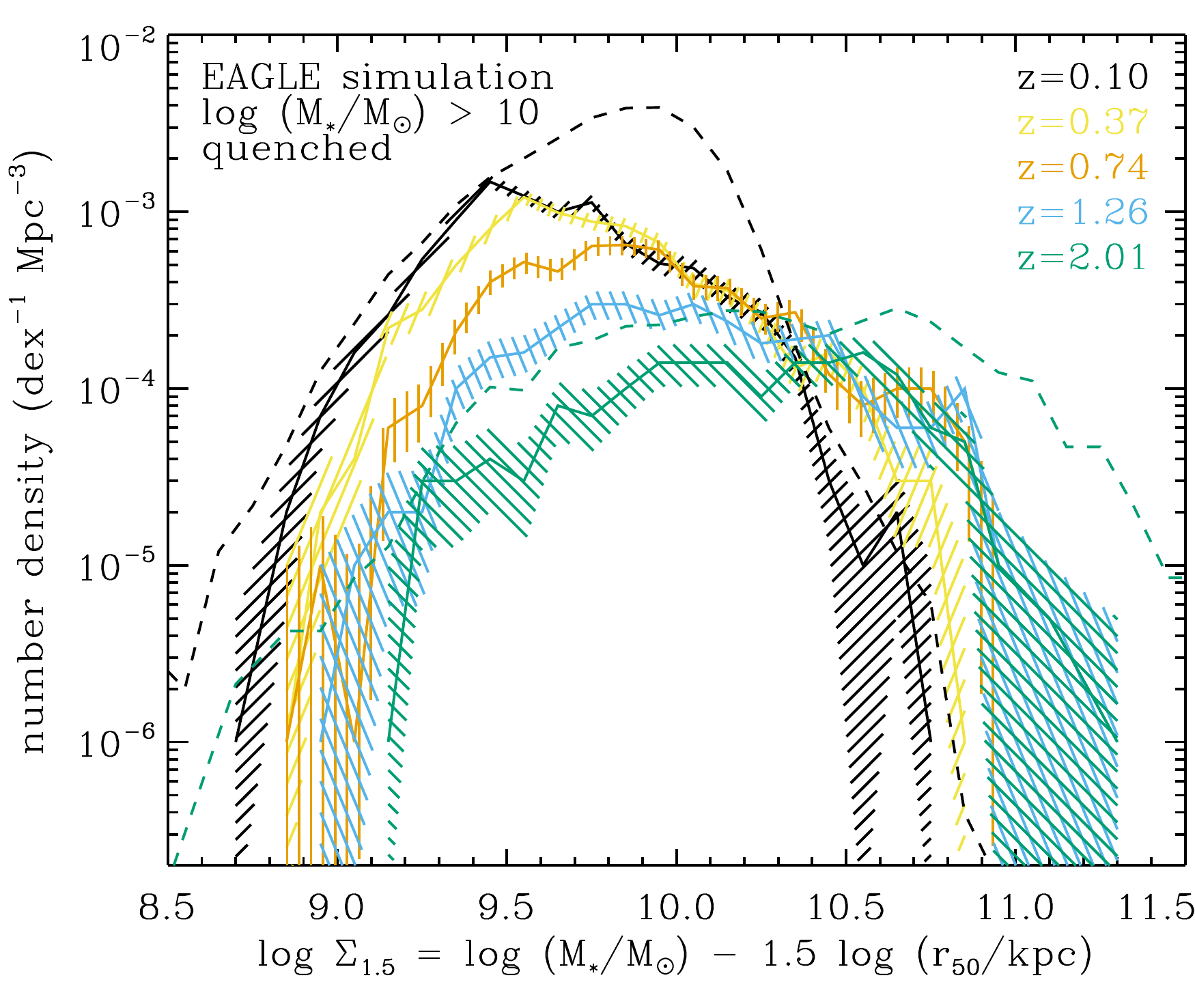}
\caption{$\log \Sigma_{1.5}$ distribution functions using the EAGLE hydrodynamical simulation.
  These are shown with solid lines, 
  while the hatched regions represent the Poisson counting uncertainties ($\pm\sigma$).
  The dashed lines show the observational measurements that are the lowest and highest redshift bins from
  the middle panel of Fig.~\ref{fig:sigma-func-evolve}.}
\label{fig:eagle-sigma-funcs}
\end{figure}

To examine the EAGLE simulation in terms of the $\log \Sigma_{1.5}$
distribution functions, we obtained data for five snapshots (different redshifts), four that
matched the midpoints of the low- and high-z comparison measurements plus one intermediate redshift,
from the $10^6\,\mathrm{cMpc}^3$ reference simulation box \citep{mcalpine16}.
Total stellar masses were defined
in spheres of radius 30\,kpc around each galaxy, with half-mass radii also defined in 3D \citep{furlong17}.
Figure~\ref{fig:eagle-sigma-funcs} 
shows the distribution functions from the simulation using these definitions of $M_{*}$ and $r_{50}$
for massive quenched galaxies using the SSFR cut of Eq.~\ref{eq:ssfr-cut}.
There are clear quantitative differences compared to the observational data: 
a lower number density at all redshifts; 
and a lack of ultra-compact galaxies at high redshift that is not unexpected given the 
gravitational softening length of 0.7 kpc \citep{schaye15}. 
There is however strikingly good qualitative agreement.
\citet{crain15} note that it is important for hydrodynamical simulations to match the observed 
size-mass relation otherwise simulations can end up with unrealistically compact low-z galaxies. 
While the EAGLE reference simulation was calibrated to the low-z relation, it was not to the 
high-z relations, and therefore this qualitative agreement is a success of the model.

\begin{figure}
\includegraphics[width=\singlecolsize\textwidth]{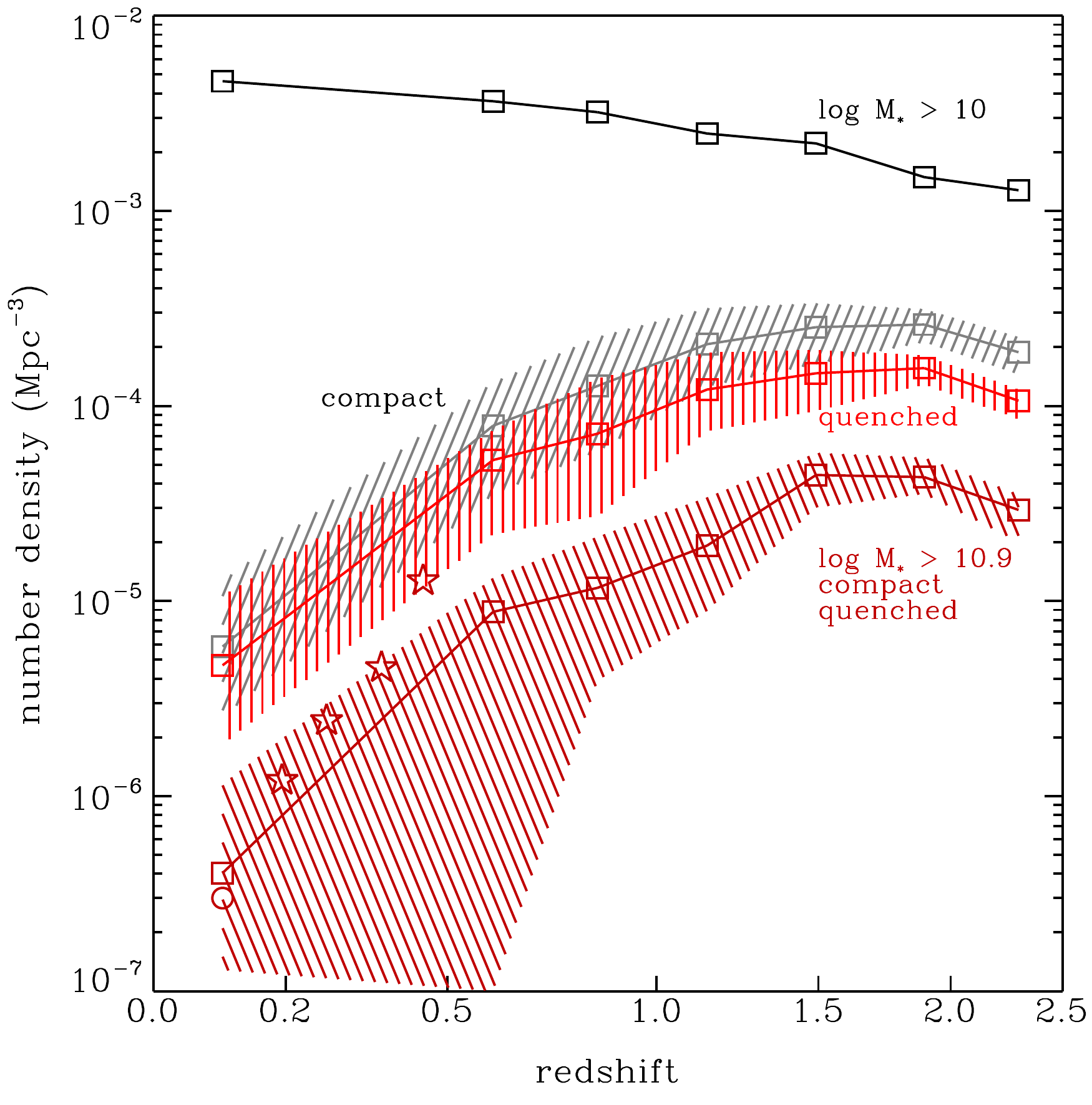}
\caption{Evolution in the number density of compact galaxies.
  The lines and squares from top to bottom represent:
  all galaxies with $\log M_{*} >10$, compact galaxies with $\log \Sigma_{1.5} > 10.5$,
  compact quenched galaxies, supermassive compact quenched galaxies.
  The hatched regions represent the range in number density from changing the definition
  of compact galaxies from $\log \Sigma_{1.5} > 10.4$ to 10.6.
\newtext{The stars show the corrected number densities for UCMGs from \citet{scognamiglio20}
  (their fig.~7, $r_{50,\mathrm{circ}} < 1.5\,\mathrm{kpc}$), while the circle shows
  the number density from the SDSS-UKIDSS sample with $\reff < 2.0\,\mathrm{kpc}$ and $\log M_{*} > 10.9$.}
}
\label{fig:numdens-compact}
\end{figure}

The distribution functions in Fig.~\ref{fig:sigma-func-evolve} clearly demonstrate 
the number density evolution without the need to define a compact sample.
Nevertheless, there is interest in the number density of compact galaxies that have 
largely evolved passively since high redshift. 
Figure~\ref{fig:numdens-compact} shows the evolution in the number density of 
compact galaxies with $\log \Sigma_{1.5} > 10.5$
(cf.\ \citealt{barro14} and \citealt{gu20} use a limit of 10.45 with circularized radii). 
To estimate the uncertainty, the number densities were also determined using cuts 
at 10.4 and 10.6. The number density of low-redshift compact galaxies depends significantly
on the cut since it is on the steep part of the $\log \Sigma_{1.5}$ distribution function,
and this is the dominant uncertainty. 
The evolutionary trend is clear though with a peak at $z \sim 2$ and a steep decline in the number
density towards low redshift \citep{cassata13,vanderwel14}. 
The $z \sim 0.1$ number density of compact quenched galaxies from this study is 
$\log (n / \mathrm{cMpc}^{-3}) = -5.3 \pm 0.4$  ($-6.4 \pm 0.5$) for $\log M_{*} > 10.0$ ($> 10.9$).

\newtext{
Different definitions of ultra-compact massive galaxies (UCMGs) have been used by other studies 
(e.g.\ \citealt{trujillo09,damjanov14,tortora16,buitrago18}). 
\citet{scognamiglio20} define UCMGs using $\log M_{*} > 10.9$ ($8 \times 10^{10} \Msun$) and
$r_{50,\mathrm{circ}} < 1.5\,\mathrm{kpc}$. 
For the UKIDSS-SDSS sample, a similar number of galaxies is obtained using $\reff < 2.0\,\mathrm{kpc}$. 
This is shown by the grey dotted box in Fig.~\ref{fig:size-mass}. 
Note that nearly all UCMGs, and all in this case, have $10.9 < \log M_{*} < 11.2$ \citep{tortora20}. 
From this cut, we obtained a number density of $3 \times 10^{-7}\mathrm{cMpc}^{-3}$ for quenched UCMGs (15 galaxies).
This is within with the upper limit at $z<0.1$ determined by \citet{buitrago18}, 
and lower than the number densities of $\sim 10^{-6}\mathrm{cMpc}^{-3}$ at $z \approx 0.2$ determined by
\citet{scognamiglio20} and of $\sim 3 \times 10^{-6}\mathrm{cMpc}^{-3}$ at $0.2 < z < 0.6$ by \citet{damjanov14}
(their fig.~7, UCMG lower mass limit), 
Our results are consistent with steep evolution in the number density at $z<0.5$ (Fig.~\ref{fig:numdens-compact}).
}

\subsection{Environment}
\label{sec:enviro}

The spectroscopic completeness of the SDSS-UKIDSS galaxy sample drops
from 92\% for normal-size galaxies to 76\% for compact galaxies (Fig.~\ref{fig:completeness}).
This is primarily due to fibre collisions. 
To evaluate the reason for and physically quantify this effect, 
we computed environmental densites for a selected galaxy sample.

A density-defining population (DDP) of galaxies was selected over an expanded
redshift range (0.024--0.168)
to avoid a redshift edge when determining densities at the sample limits of 0.04 and 0.15.
The DDP was selected: 
(i) with $\log M_{*} > 10.5$ or $\log M_{*} > 10.0$ for quenched galaxies; and
(ii) with a spec-z or a photo-z that had a nominal error $\zeta_{i,\mathrm{err}}^2  < 0.015$.
For a calibration sample with both spec-z and good photo-z (within this error limit),
90\% had $\Delta \ln (1+z) < 0.015$ (difference between the spec-z and photo-z).
This DDP thus supplements the spec-z with accurate photo-z enabled by the SDSS-UKIDSS data 
and significantly reduces biases associated with fibre collisions when measuring environmental densities. 

For each galaxy, the projected distances to the three nearest DDP neighbours were determined.
The neighbours had to have $\Delta v < 1200$\,km/s using spec-z or
$\Delta v < 4500$\,km/s using photo-z (for either the potential neighbour and/or the target), 
where $\Delta v = c \Delta \ln(1+z)$ is the velocity difference \citep{Baldry18zeta}
between the target and DDP galaxy.
The surface neighbour density for each galaxy is then given by
\begin{equation}
  \log \Sigma_{{\rm ddp}} = \frac{1}{2} \log \left( \frac{2}{\pi d_2^2} \right) +
                          \frac{1}{2} \log \left( \frac{3}{\pi d_3^2} \right) 
\end{equation}
where $d_2$ and $d_3$ are the projected comoving distances (cMpc) to the 2nd and 3rd nearest neighbours.
This is then converted to an environmental overdensity ($\delta$) using
\begin{equation}
  \log (1+ \delta) = \log \Sigma_{{\rm ddp}} - f_{\rm ddp}(z)  \mbox{~~~,}
\end{equation}
where $f_{\rm ddp}(z)$ is a fitted function for the global surface density of the DDP
(in bins of 2400\,km/s)
that accounts for the observational flux limit because the DDP is not a volume-limited sample.
For this DDP and $r_{\rm petro} < 17.8$, $f_{\rm ddp}(z)$ was approximated by 
$f_{\rm ddp} = -0.978$ for $z \le 0.0784$ and $-0.351 - 8.30 \ln(1+z)$ otherwise ($-1.511$ at $z=0.15$).
This ensures that the median value of $\delta$ is similar across the full redshift range 
for a particular type of galaxy. 
This was tested for both quenched and star-forming galaxies with $\log M_{*} > 10.5$. 

\begin{figure}
\includegraphics[width=\singlecolsize\textwidth]{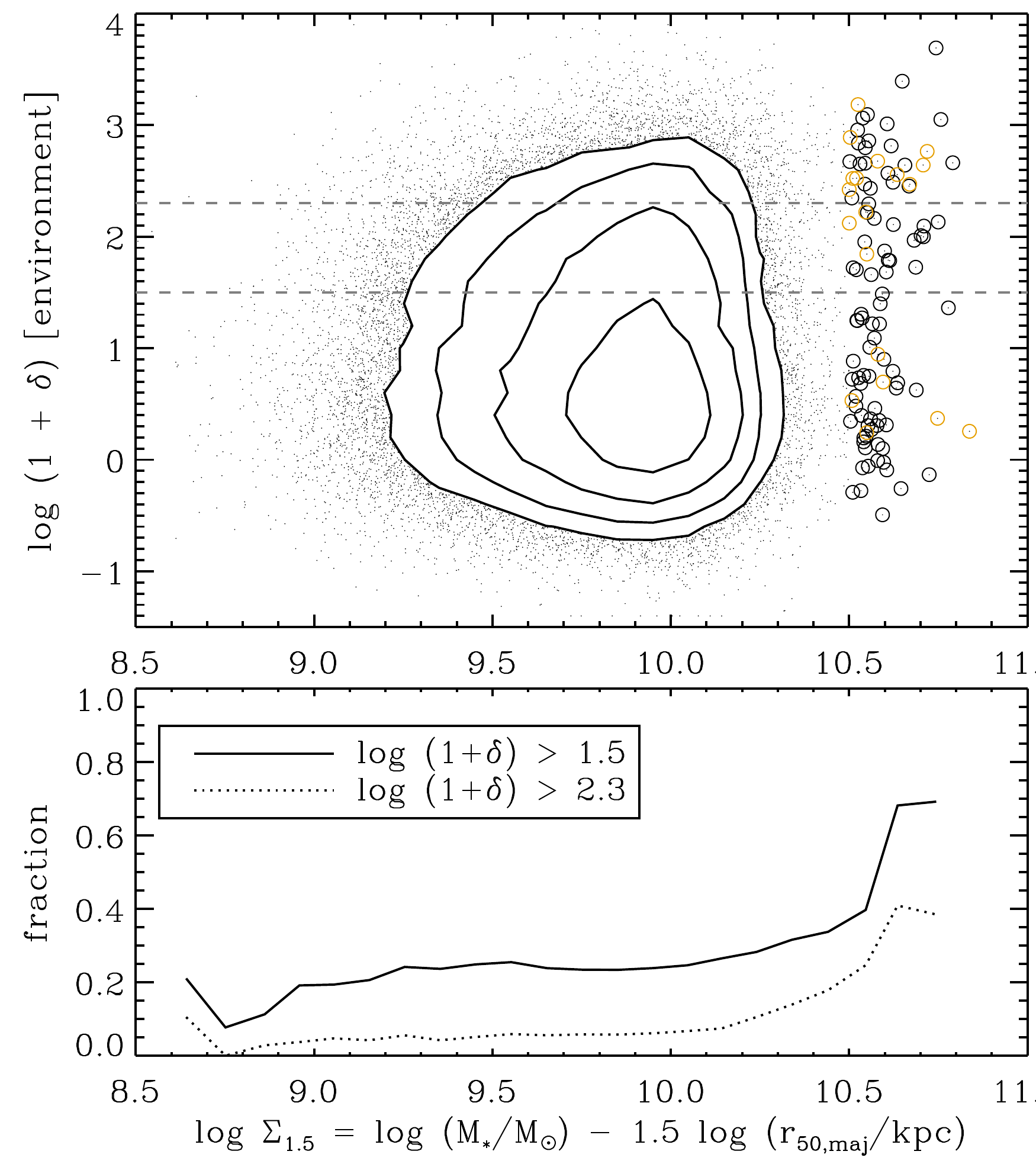}
\caption{Upper panel: environmental densities for the quenched
  galaxy sample ($0.04 < z < 0.15$). Lower panel: fraction of galaxies
  in high density environments. 
  The circles represent galaxies with $\log \Sigma_{1.5} > 10.5$ 
  (black with spec-z and orange with photo-z)}
\label{fig:environment}
\end{figure}

Figure~\ref{fig:environment} shows the environmental overdensity versus $\log \Sigma_{1.5}$
for the quenched galaxies. The lower panel shows the fraction of galaxies in high density
environments with $\log (1+\delta) > 1.5$. 
From this, there is a clear difference with $69 \pm 8$\% of the most compact galaxies ($>10.6$) found in
high density environments compared to 24\% for normal-size quenched galaxies.
This noticeable effect is in contrast to the median $\log \Sigma_{1.5}$ that is 9.875 in high-density
environments compared to 9.866 at lower densities. In other words, for quenched galaxies,
there are small or negligible differences between the general size-mass relations at low redshift 
\citep{maltby10,matharu19}.
\citeauthor{matharu19} noted that while minor mergers in clusters are less likely than in the field, 
the dissapearance of compact cluster galaxies that are seen at high redshift can be explained 
by a combination of mergers with the brightest cluster galaxy and being tidally destroyed. 
For the rarer compact galaxies, they are nevertheless more likely to survive 
in high-density environments \citep{poggianti13,trujillo14,buitrago18}. 
This is demonstrated conclusively using this complete SDSS-UKIDSS sample
\newtext{(see Fig.~\ref{fig:images2} for images of compact galaxies in high-density environments).} 


\newtext{
  This result is consistent with analysis of \citet{peralta16}, 
  using an SDSS sample, who showed that the fraction of relics (compact early types)
  relative to normal-sized early types was higher in high-density environments.
  For our sample, we find that 0.7\% of quenched massive galaxies are compact ($\log \Sigma_{1.5} > 10.5$) in 
  high density environments ($\log (1+\delta) > 2.3$) compared to 0.1\% in low density ($\log (1+\delta) < 1.5$).
  For $\log \Sigma_{1.5} > 10.2$, we find 7.8\% compared to 3.7\%.
  \citet{damjanov15} found a neglible difference between the environments of compact and non-compact quiescent
  galaxies when controlling for stellar mass (their fig.~10) at $0.1 < z < 0.4$.
  \citet{tortora20} found UCMGs to be slightly more abundant in clusters compared to the field when restricting
  the comparison sample to $10.9 < \log M_{*} < 11.2$ at $z \sim 0.4$ (their fig.~1). 
  These differences compared to the SDSS-UKIDSS sample
  could arise from the types of density measurements, the mass range or from evolution since $z \sim 0.4$. 
}

\section{Summary}
\label{sec:summary}

The primary aim of this analysis was to determine the number density of compact galaxies using a complete
sample with photometry from SDSS and UKIDSS.
After masking around GAIA stars (Figs.~\ref{fig:exclusion-examples}--\ref{fig:exclusion-radius-fit}),
rather than using a not-saturated criterion,
and selecting galaxies using primarily colours ($JKgi$, $YKgz$,
Figs.~\ref{fig:color-color}--\ref{fig:delta-color}),
a sample of 245\,645 sources was obtained covering $2300\,\sqdeg$ to $r < 17.8$.
The majority (92\%) of the sample were selected without using a profile separator,
i.e.\ with no bias against compact galaxies, 
with the additional selection being mainly extended sources
that did not have catalog-matched UKIDSS photometry and which were added in for completeness. 

To refine the sample further, photo-z were determined using an accurate scaled flux matching (SFM)
empirical method. The matching set were galaxies with reliable spec-z and photometry.
A photo-z estimate can be obtained for any source
even if it is missing or has bad photometry in one or more bands.
The criteria for a reliable photo-z estimate were determined using:
the number of matches within a chi-squared limit (Fig.~\ref{fig:count-chisq}),
and the estimate of the error (Fig.~\ref{fig:pz-err}).
For sources with reliable spec-z and photo-z,
the 68th percentile of $|\Delta \ln(1+z)|$ was 0.0125
(0.010 for quenched galaxies, 0.016 for star-forming galaxies)
and 99.7\% were within 0.06 (Fig.~\ref{fig:spec-z-photo-z}). 
These photo-z were more accurate than the SDSS \citet{beck16} photo-z
because of the additional near-IR photometry.

Sources were assigned reliable photo-z if reliable spec-z were not available; 
otherwise they were not included and assumed to be stars or quasars. 
The selection of sources with $0.04 < z < 0.15$ produced a sample 
of 163\,186 galaxies (10.4\% with photo-z, Fig.~\ref{fig:regions}). 
Utilizing SDSS pipeline half-light radii that were shown to be similar to
the sizes from the two-component fits of \citet{simard11} (Fig.~\ref{fig:size-comparison}),
a size-mass distribution was produced (Fig.~\ref{fig:size-mass}).
This is complete for compact galaxies because of the colour selection and photo-z.
The slope of the high-mass relation for quiescent galaxies is is about 2/3,
and thus following \citet{barro13},
we computed $\Sigma_{1.5}$
($M_{*}/\reff^{1.5}$ in $\Msun\mathrm{kpc}^{-1.5}$, Eq.~\ref{eq:sigma-1pt5}) that runs perpendicular
to this slope.
The spectroscopic completeness drops from 92\% for normal-size galaxies to 76\% for compact galaxies
with $\log \Sigma_{1.5} > 10.5$ (Fig.~\ref{fig:completeness}) ($\log M_{*} > 10$).
This can be attributed primarily to fibre collisions, and thus environments, 
rather than incompleteness of the SDSS target selection.


A high-z comparison sample was obtained from 3D-HST data with 
structural measurements from \citet{vanderwel12} and stellar masses from \citet{driver18}. 
The evolution in the $\log \Sigma_{1.5}$ distribution functions (Fig.\ref{fig:sigma-func-evolve}) 
from high to low redshift shows: a gradual rise in the number density at $\log \Sigma_{1.5} \la 10$; 
while at $> 10$, there is a sharp cutoff in the number density at lower redshifts.
In the SDSS-UKIDSS sample, the number densities of compact galaxies are about a factor of 30--100 
below the peak at $z \sim 2$ (Fig.~\ref{fig:numdens-compact}).
The consensus mechanism for the growth of compact galaxies is primarily via minor mergers. 

Comparing the distribution functions with different redshift snapshots from the
EAGLE hydrodynamical simulation shows good qualitative agreement (Fig.~\ref{fig:eagle-sigma-funcs}).
In addition to the GSMF (Fig.~\ref{fig:gsmf-evolve}) and size-mass relations, 
using $\log \Sigma_{1.5}$ distributions is an informative way to compare simulations with data
because the number densities for both normal-size and compact galaxies are clearly conveyed. 
This is an avenue for future research, for example, for simulations that agree with observations,
the past merger history and environments 
of quenched galaxies with different low- and high-redshift $\log \Sigma_{1.5}$ values could be compared. 

We searched amongst the entire population of (more than two million) stars and galaxies to $r<17.8$,
using 9-band photometry, for compact galaxies.
This confirmed the low number densities ($\la 10^{-5}\,\mathrm{cMpc}^{-3}$)
of local compact galaxies in agreement with the SDSS analysis of \citet{taylor10}. 
The $\log \Sigma_{1.5}$ distribution of local quenched galaxies is similar
in low- and high-density environments (Fig.~\ref{fig:environment} upper panel).
However, the rare compact galaxies are significantly more likely to survive in
high density environments (Fig.~\ref{fig:environment} lower panel).

There is substantial evolution in the size-mass distribution since $z \sim 1$.
In the future, the WAVES-DEEP survey with 4MOST aims to obtain redshifts 
for galaxies to $z \la 0.8$ over $\sim 60\,\sqdeg$ \citep{WAVES19}.
Coupled with Euclid imaging \citep{laureijs10,dasilva19},
this will provide galaxy structural measurements over a cosmic volume that is sufficient
to divide the sample into several epochs and different types of environments.
The diversification of the compact galaxy population will be illuminated in detail. 

\section*{Data Availability}

The data underlying this article were accessed from: 
the Sloan Digital Sky Survey at skyserver.sdss.org/ (dr14.PhotoPrimary, dr14.SpecObj, dr14.Photoz, dr7.PhotoObjAll); 
the WFCAM Science Archive at wsa.roe.ac.uk/ for the UKIDSS data (lasYJHKsource); 
the Gaia archive at gea.esac.esa.int/archive/ (gaiadr2.gaia\_source); 
and the 3D-HST archive at 3dhst.research.yale.edu/
[the catalog is also available via ADS 2015yCat..22140024S \citep{skelton14}]. 
Galaxy structural measurements can be accessed via
ADS 2011yCat..21960011S \citep{simard11} and ADS 2012yCat..22030024V \citep{vanderwel12}. 

The derived data generated in this research will be shared at 
{\tt www.astro.ljmu.ac.uk/\midtilde{}ikb/research/}
or on reasonable request to the corresponding author.

\section*{Acknowledgements}

We thank
Simon Driver for providing the stellar masses,
Arjen van der Wel for providing structural measurements for the high-z comparison sample, 
Crescenzo Tortora for providing structural measurements from the KiDS imaging, 
and the anonymous referee for careful and useful comments. 

Funding for the SDSS and SDSS-II has been provided by the Alfred P. Sloan Foundation,
the Participating Institutions, the National Science Foundation,
the U.S. Department of Energy, the National Aeronautics and Space Administration,
the Japanese Monbukagakusho, the Max Planck Society, and the Higher Education Funding Council for England.

This work is based in part on data obtained as part of the UKIRT Infrared Deep Sky Survey.

This work is based in part on observations taken by the 3D-HST Treasury Program (GO 12177 and 12328) with the NASA/ESA HST, which is operated by the Association of Universities for Research in Astronomy, Inc., under NASA contract NAS5-26555. 

We acknowledge the Virgo Consortium for making their simulation data available.
The EAGLE simulations were performed using the DiRAC-2 facility at Durham, managed by the ICC,
and the PRACE facility Curie based in France at TGCC, CEA, Bruyeres-le-Chatel.

\setlength{\bibhang}{2.0em}
\setlength\labelwidth{0.0em}

\bibliographystyle{mn2e-williams}
\bibliography{galaxies,surveys,stars,general,cosmology}

\appendix

\section{Size tests and Images}

This paper uses the SDSS pipeline profile fits for the estimates of the
half-light radii (\S~\ref{sec:half-light}).
The photometric pipeline was constructed by \citet{lupton01} and 
is described primarily in \S~4.4 of \citet{stoughton02} with updates in other 
data release papers (e.g.\ \citealt{sdssDR2}). 
The measurements are robust because the de Vaucouleurs (Sersic $n=4$) and 
exponential (Sersic $n=1$) profiles are fitted separately.
Therefore, minimization problems are less severe 
compared to simultaneous bulge-plus-disk fitting 
or allowing for a free Sersic index (e.g.\ \citealt{simard11,kelvin12}).
In addition, the local PSF is well determined and accounts 
for the variation across each detector chip (\S~4.3 of \citeauthor{stoughton02}).
This variation is only 1-dimensional because SDSS images were obtained using
drift scanning.
This is a notable advantage for empirical PSF determination compared to
typical point-and-stare integrations.

\begin{figure}
  \centerline{\includegraphics[width=\singlecolsize\textwidth]{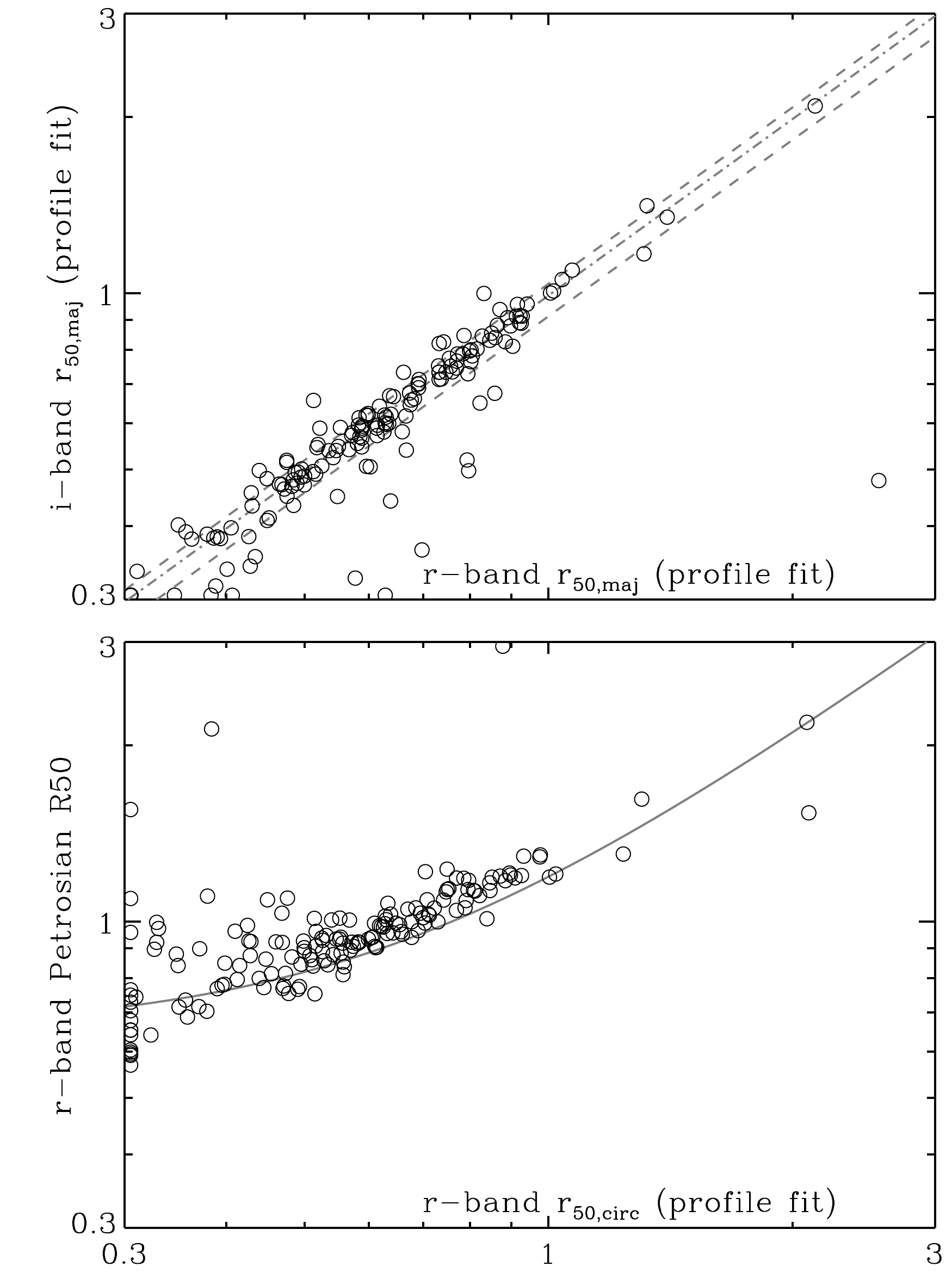}}
  \caption{Internal size tests for the compact galaxies with $\log \Sigma_{1.5} > 10.5$. Top panel:
    comparison between $r$- and $i$-band measurements from the profile fits (units of arcsec).
    The dot-dash line shows the median ratio, while the dashed lines show the 16th and 84th percentiles.
    Bottom panel: comparison between the Petrosian half-light radii and the profile fits.
    The solid line shows $y = \sqrt (x^2 + 0.65^2)$ as a simple guide to account approximately
    for the effect of the PSF on the \textsc{petroR50} sizes.}
  \label{fig:size-test}
\end{figure}

For the half-light radii, we used a weighted average of the profile radii
for each galaxy (Eq.~\ref{eqn:half-light}).
Figure~\ref{fig:size-test} (top) shows a comparison between $r$- and $i$-band $\reff$ measurements
for compact galaxies. For these galaxies, the median ratio is 0.99 ($i$- divide $r$-band size) and there is good
agreement between the measurements with 93\% differing by less than 0.1\,dex.
An alternative measure of half-light radius obtained by the SDSS pipeline is using the Petrosian
magnitude with circular apertures (\textsc{petroR50}, \citealt{stoughton02}). Figure~\ref{fig:size-test} (bottom)
shows \textsc{petroR50} versus the circularized half-light radii from the profile fits. 
The \textsc{petroR50} measurement includes the effects of the PSF. To account for this,
the solid shows a simple correction with 0.65 added in quadrature to x-axis value, where
$0.65''$ is median value of \textsc{petroR50} for stars in the SDSS-UKIDSS sample.
While most of the \textsc{petroR50} values lie above this line, 90\% are within 0.1\,dex.
The profile-fit radii will be more accurate as they account for the variable PSF.
There is sufficient agreement to demonstrate the internal consistency of the SDSS
half-light radii. 

\begin{figure}
  \centerline{\includegraphics[width=\singlecolsize\textwidth]{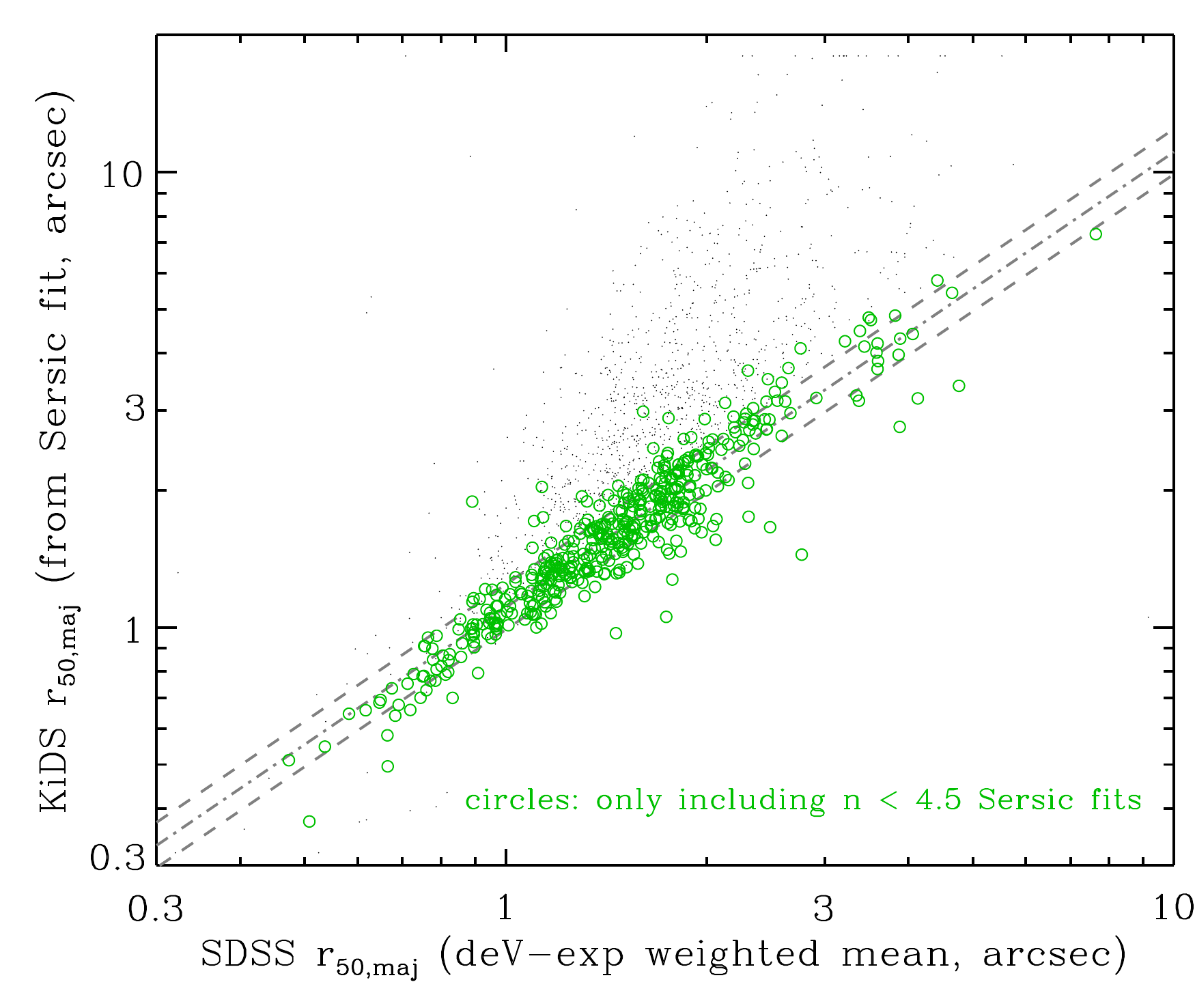}}
  \caption{Comparison between half-light radii from KiDS with SDSS.
    The dot-dash line shows the median ratio, while the dashed lines show the 16th and 84th percentiles
    for the sample shown by green circles.}
  \label{fig:size-test2}
\end{figure}

To compare with an external dataset, we used the $r$-band structural parameters determined by \citet{roy18}, 
and presented in \citet{tortora18}, based on the imaging from the Kilo Degree Survey (KiDS, \citealt{dejong17}).
The mean PSF full width half maximum (FWHM) is $0.7''$ and thus higher resolution than the SDSS which has a median of $1.3''$.
The comparison set of galaxies was expanded to $\log \Sigma_{1.5} > 10$ (using the smaller $\reff$ from KiDS or SDSS)
and $0.04 < z < 0.2$ because the overlap area was only about $200\,\sqdeg$ to give a sample of 1674 galaxies. 
Figure~\ref{fig:size-test2} shows the comparison between the half-light radii determined from KiDS
(that used a Sersic fit) with SDSS. Notably, there are many outliers shown by the points, however,
these occur primarily when the Sersic index $n$ is above 4.5. A unrealistic fraction of galaxies (72\%)
are fitted with $n>4.5$ in the KiDS process. Restricting the sample to only those galaxies where the KiDS process
fitted the galaxies with $n<4.5$ (469 galaxies), shown by the green circles, results in good agreement between the KiDS and
SDSS sizes. For these, the median offset of the KiDS sizes is $+0.04$\,dex and 84\% differ by less than 0.1\,dex.
We conclude that the SDSS sizes are more robust that the sizes determined from KiDS with a free Sersic index,
with sufficient agreement with $n<4.5$ to demonstrate that the SDSS sizes are not substantially underestimating the
half-light radii.

\begin{figure*}
 10.0
 \includegraphics{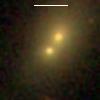}
 \includegraphics{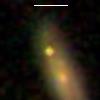}
 \includegraphics{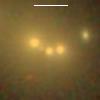}
 \includegraphics{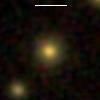}
 \includegraphics{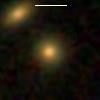}
 \includegraphics{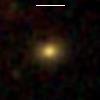}\\
 10.4
 \includegraphics{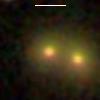}
 \includegraphics{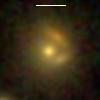}
 \includegraphics{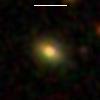}
 \includegraphics{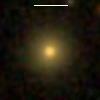}
 \includegraphics{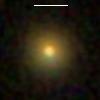}
 \includegraphics{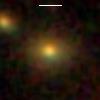}\\
 10.6
 \includegraphics{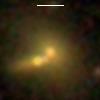}
 \includegraphics{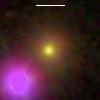}
 \includegraphics{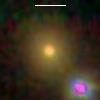}
 \includegraphics{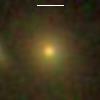}
 \includegraphics{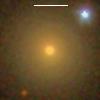}
 \includegraphics{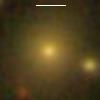}\\
 10.7
 \includegraphics{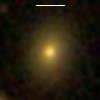}
 \includegraphics{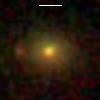}
 \includegraphics{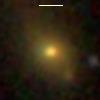}
 \includegraphics{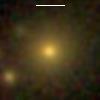}
 \includegraphics{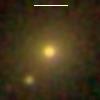}
 \includegraphics{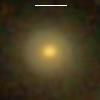}\\
 10.8
 \includegraphics{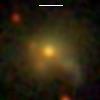}
 \includegraphics{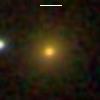}
 \includegraphics{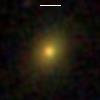}
 \includegraphics{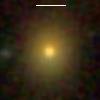}
 \includegraphics{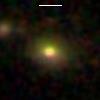}
 \includegraphics{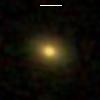}\\
 11.0
 \includegraphics{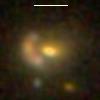}
 \includegraphics{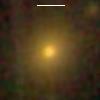}
 \includegraphics{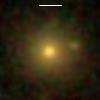}
 \includegraphics{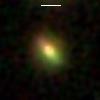}
 \includegraphics{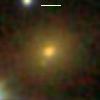}
 \includegraphics{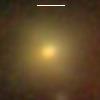}\\
 10.3
 \includegraphics{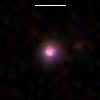}
 \includegraphics{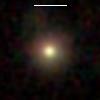}
 \includegraphics{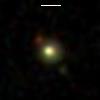}
 \includegraphics{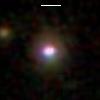}
 \includegraphics{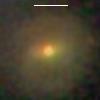}
 \includegraphics{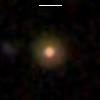}\\
 10.6
 \includegraphics{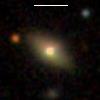}
 \includegraphics{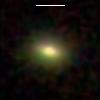}
 \includegraphics{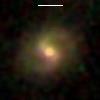}
 \includegraphics{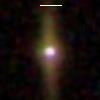}
 \includegraphics{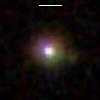}
 \includegraphics{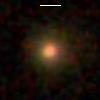}
 \caption{Images of compact galaxies ($\log \Sigma_{1.5} > 10.5$). Most images are $20'' \times 20''$;
   ten images centred on galaxies at $z<0.08$ were scaled to $30 \times 30\,\mathrm{kpc}$.
   The white bar at the top of each image represents 10\,kpc at the redshift of the central galaxy. 
   The top six rows show the quenched the population, 
   while the remaining two rows show galaxies with higher star-formation rates. 
   The images are ordered from the lowest to highest in stellar mass (10.0--11.4) within each subsample,
   and are evenly selected by mass from the 164 shown with circles in Fig.~\ref{fig:size-mass}.
   The mass ($\log M_{*}$) is noted for the first image in each row. 
   The arcsinh stretch used to produce the images ($i$-$r$-$g\rightarrow$RGB) is described by \citet{lupton04}.}
 \label{fig:images1}
\end{figure*}

The final check on the compact galaxies is visual inspection, 
noting that those in areas of severe scattered light, not accounted for by the GAIA masking,
have been rejected from the sample. 
Images were obtained from the SDSS cutout service \citep{NSG04} scaled as per \citet{lupton04}. 
Figure~\ref{fig:images1} shows a selection of 48 out of the 164 compact galaxies.
Each image clearly shows a concentration of light within a small area ($\sim 2''$)
with the size of the core being determined by the PSF in many cases.
By chance, the first image in row six shows a possible lense of a background galaxy
(coordinates 343.09313, $+01.11666$; \citealt{diehl17}).

Note that a number of images show an obvious extended halo around the compact core. 
The use of the arcsinh stretch \citep{lupton04} is specifically designed to highlight low-surface-brightness features 
at the same time as preserving the structure of bright sources.
Nevertheless, an extended halo may indicate that the SDSS pipeline half-light radii apply to the core and not to the whole galaxy.
Taking a sample of 111 compact galaxies that are quenched ($\log \Sigma_{1.5} > 10.5$ and $\reff < 2$), 
we tested this by using the \textsc{PhotoProfile} measurements from SDSS to determine $r$-band aperture magnitudes
out to radii of 5\,kpc and 20\,kpc (total flux for this test) where available.
From these and by visual inspection (to rule out problems with the aperture magnitudes), 
only about 15 out of 111 have $>10$\% of their total flux at radii larger than 5\,kpc, with the median being 3\%. 
For 40 star-forming compact galaxies, the median is 13\% of the total flux at $r>5\,\mathrm{kpc}$ in the $r$-band. 
For all 164 galaxies with $\log \Sigma_{1.5} > 10.5$, these total aperture magnitudes differ from the \textsc{cmodel} magnitudes
by less than 0.10\,mag for 150 galaxies, and by less than 0.15\,mag for 159 galaxies.
This demonstrates that the SDSS pipeline profile fits are accounting for most of the flux, 
and are thus sufficiently representative of each galaxy for determining the half-light radii and masses. 

\begin{figure*}
 \includegraphics{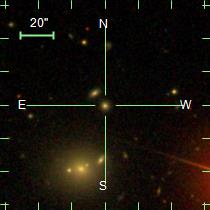}
 \includegraphics{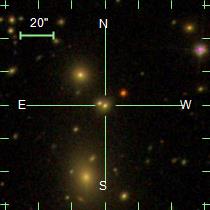}
 \includegraphics{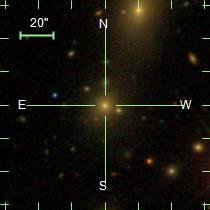}
 \includegraphics{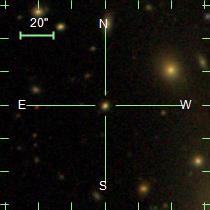}
 \includegraphics{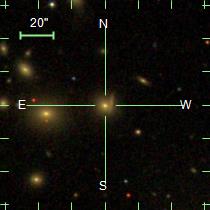}
 \includegraphics{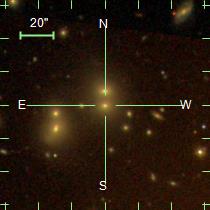}
 \includegraphics{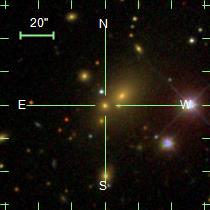}
 \includegraphics{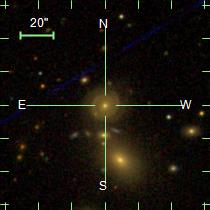}
 \includegraphics{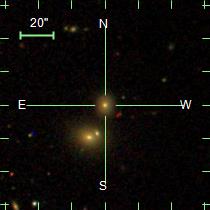}
 \includegraphics{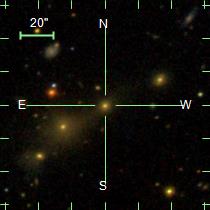}
 \includegraphics{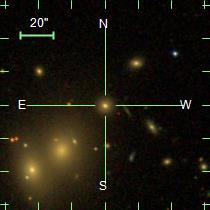}
 \includegraphics{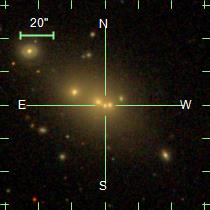}
 \caption{Images centred on compact galaxies in high-density environments. 
   The second image has two compact galaxies. The images are selected 
   from the compact galaxies with $\log (1+\delta) > 2.3$ shown in Fig.~\ref{fig:environment}.}
 \label{fig:images2}
\end{figure*}

Figure~\ref{fig:images2} shows a larger sky area around a sample of the compact galaxies
in high-density regions.
These images demonstrate the fidelity of the compact galaxy sample. 

\label{lastpage}

\end{document}